\documentclass[10pt]{article}

\usepackage{amsmath,amstext,amsgen,latexsym}
\usepackage{amstext,amssymb,amsfonts,latexsym}
\usepackage{theorem}
\usepackage{pifont}
\usepackage{graphics,epsfig}

 \newcommand{\bs}{\bigskip}
 \newcommand{\ms}{\medskip}
 \newcommand{\n}{\noindent}
 
 \newcommand{\hs}[1]{\hspace*{ #1 mm}}
 \newcommand{\vs}[1]{\vspace*{ #1 mm}}

 \newcommand{\forces}{\parallel\hspace*{-2mm}-\,}
 \newcommand{\setempty}{\mathrm{\O}}

 \newcommand{\nat}{\mathbb{N}}
 
 \newcommand{\integer}{\mathbb{Z}}

 \newcommand{\dom}{\mbox{dom}}
 
 \newcommand{\ran}{\mbox{ran}}

 \newcommand{\co}{\mathrm{co}\mbox{-}}

 \newcommand{\ie}{\textrm{i.e.},\hspace*{2mm}}
 \newcommand{\eg}{\textrm{e.g.},\hspace*{2mm}}
 
 \newcommand{\etalc}{\textrm{et al.}}

 \newcommand{\AAA}{{\cal A}}

 \newcommand{\CC}{{\cal C}}
 \newcommand{\FF}{{\cal F}}
 \newcommand{\DD}{{\cal D}}

 \newcommand{\p}{\mathrm{P}}
 \newcommand{\np}{\mathrm{NP}}
 \newcommand{\ph}{\mathrm{PH}}

 \newcommand{\up}{\mathrm{UP}}

 \newcommand{\e}{\mathrm{E}}
 \newcommand{\nde}{\mathrm{NE}}
 
 \newcommand{\dexp}{\mathrm{EXP}}
 
 \newcommand{\bh}{\mathrm{BH}}

 \newcommand{\subexp}{\mathrm{SUBEXP}}

 \newcommand{\cequalp}{\mathrm{C}_{=}\mathrm{P}}

 \newcommand{\fp}{\mathrm{FP}}

 \newcommand{\npsv}{\mathrm{NPSV}}

 \newcommand{\unique}{\mathrm{U}}

 \newcommand{\deltah}[1]{\Delta^{\mathrm{P}}_{#1}}
 \newcommand{\sigmah}[1]{\Sigma^{\mathrm{P}}_{#1}}
 \newcommand{\pih}[1]{\Pi^{\mathrm{P}}_{#1}}

 \newcommand{\deltaexph}[1]{\Delta^{\mathrm{EXP}}_{#1}}

 \newcommand{\fdeltah}[1]{\mathrm{F}\Delta^{\mathrm{P}}_{#1}}
 \newcommand{\subdeltaexph}[1]{\mathrm{SUB}\Delta^{\mathrm{EXP}}_{#1}}

 \newcommand{\Treduces}[1]{\leq_{T}^{#1}}

 \newcommand{\onereduces}[1]{\leq_{1}^{#1}}

 \newcommand{\honettreduces}[1]{\leq_{1tt}^{h\mbox{-}#1}}

 \newcommand{\IFF}{\Longleftrightarrow}

 \def\bbox{\vrule height6pt width6pt depth1pt}

\theoremstyle{plain}
\theoremheaderfont{\bfseries}
 \newtheorem{theorem}{Theorem}[section]
 \newtheorem{lemma}[theorem]{Lemma}
 \newtheorem{proposition}[theorem]{Proposition}
 \newtheorem{corollary}[theorem]{Corollary}

 {\theorembodyfont{\rmfamily}
\newtheorem{definition}[theorem]{Definition}}
 {\theorembodyfont{\rmfamily} }
 {\theorembodyfont{\rmfamily} }

 \newenvironment{claim}{\vspace*{3mm} \par \noindent
            {\bf Claim. \hs{1}}}{\vspace*{3mm}}

 \newenvironment{proof}{\par \noindent
            {\bf Proof. \hs{2}}}{\hfill$\Box$ \vspace*{3mm}}

 \newenvironment{proofof}[1]{\vspace*{5mm} \par \noindent
         {\bf Proof of #1.\hs{2}}}{\hfill$\Box$ \vspace*{3mm}}

 \newcommand{\floors}[1]{\lfloor #1 \rfloor}
 \newcommand{\pair}[1]{\langle #1 \rangle}

 \newcommand{\dtime}[1]{{\mathrm{DTIME}}(#1)}

 \newcommand{\ignore}[1]{}
\newif\ifnotesw\noteswtrue
   {\ifnotesw\marginpar[\hfill\(\top\)]{\(\top\)}\fi}%
      {\ifnotesw\marginpar[\hfill\(\bot\)]{\(\bot\)}\fi}
      
\newcommand{\mnote}[1]%
   {\ifnotesw\marginpar%
	  [{\scriptsize\begin{minipage}[t]{\marginparwidth}
	  \raggedleft#1%
		  \end{minipage}}]%
	  {\scriptsize\begin{minipage}[t]{\marginparwidth}
	  \raggedright#1%
		  \end{minipage}}%
    \fi}
 \newcommand{\marked}{\mathit{MARKED}}
 \newcommand{\gindex}{\mathit{INDEX}}
\begin{document} 
\pagestyle{plain}
\setcounter{page}{1}
\begin{center}
{\Large {\bf Resource Bounded Immunity and Simplicity}}
\footnote{This paper partly extends the unpublished manuscript of the first author \cite{Yam95}. A preliminary version appeared in the Proceedings of the 3rd IFIP International Conference on Theoretical Computer Science, Kluwer Academic Publishers, pp.81--95, Toulouse, France, August 23--26, 2004. This work was in part done while the second author served as a postdoctoral fellow at the University of Ottawa between September and December of 2000. This work was in part supported by the Natural Sciences and Engineering Research
Council of Canada and Grant-in-Aid for Scientific Research 
(No. 14740082), Japan Ministry of Education, Culture, Sports, Science, and Technology.}

\bs

\begin{tabular}{c@{\hspace{20mm}}c}
{\sc Tomoyuki Yamakami}${}^{\small a}$ & {\sc Toshio Suzuki}${}^{\small b}$  
\end{tabular}

\ms

${}^{\small a}${\small {\em School of Information Technology and Engineering, University of Ottawa, Ottawa, Ontario, Canada K1N 6N5}} \\
${}^{\small b}${\small {\em Department of Mathematics and Information Sciences, Osaka Prefecture University, Osaka, 599-8531  Japan}} 
\end{center}

\paragraph{Abstract:}
Revisiting the thirty years-old notions of resource-bounded immunity and simplicity, we investigate the structural characteristics of various immunity notions: strong immunity, almost immunity, and hyperimmunity as well as their corresponding simplicity notions. We also study limited immunity and simplicity, called $k$-immunity and feasible $k$-immunity, and their simplicity notions. Finally, we propose the k-immune hypothesis as a working hypothesis that guarantees the existence of simple sets in $\np$. 

\bs

\n{\sf Key Words:}\hs{2} immune set, 
simple set, complete set, weak forcing, generic oracle, random oracle

\section{Prologue}\label{sec:introduction}

The twentieth century opened its curtain with a keynote speech of Hilbert on the list of open problems that should be challenged in the coming century. One problem of his relates to a fundamental question on the computability. In his speech, Hilbert raised the question, known as Hilbert's tenth problem, of whether there exists a ``procedure'' to calculate integer solutions of polynomial equations. This question signifies the importance of {\lq\lq}algorithmic procedure.{\rq\rq} Subsequently, the twentieth century had delivered the new theory of computation and computability. The notion of computability was first introduced by G{\"o}del \cite{God31} in his epochal paper on the incompleteness of any consistent theory. Matijasevich's solution to Hilbert's tenth problem is a remarkable success of classical recursion theory.

In the last half century, a new breed of computation theory---computational complexity theory---hatched from classical recursion theory as electronic computing device has been materialized. In particular, the classes $\p$ and $\np$ are defined analogously to the classes of recursive sets and recursively enumerable (r.e., in short) sets, respectively.  By the pioneering work of Cook, Karp, and Levin, the theory of {\lq\lq}$\np$-completeness{\rq\rq} (more precisely, $\p$-m-completeness for $\np$) has boosted the progress of complexity theory. Since the 1970s, the complexity class $\np$ has been widely recognized as an important complexity class that includes many natural problems in computer science, and structural complexity theory has evolved revolving around this class $\np$. A glossary of $\p$-m-complete problems for $\np$ are found in, for instance, the textbook of Gary and Johnson \cite{GJ79}. In the early 1970s, Meyer and Stockmeyer \cite{MS73} further defined the {\em polynomial (time) hierarchy}, which is built over $\np$ by a way of relativization, and this notion was refined later by Stockmeyer \cite{Sto77} and Wrathall \cite{Wra77}. It is strongly believed by a number of complexity theoreticians that $\np$ differs from $\p$ and, moreover, the polynomial hierarchy indeed forms an infinite hierarchy. If $\np$ properly includes $\p$, as is believed, then how hard would $\np$ be? As Ladner \cite{Lad75} demonstrated, assuming that $\p\neq\np$, there are an infinite number of layers of polynomial-time equivalence classes in the gap between $\p$ and the class of $\p$-m-complete sets for $\np$. Is there any common characteristic among $\np$ sets that do not fall into $\p$? Sch{\"o}ning's \cite{Sch83} low and high hierarchies within $\np$, for instance, partially fill the difference $\np\setminus \p$. It is thus of great importance to study the structural characteristics of the sets sitting in this difference.  

Back in the 1940s, Post \cite{Pos44} asked whether there exists an intermediate {r.e.\!\!} set that is neither recursive nor Turing-complete. As an attempt to answer this question, he constructed a non-recursive set, called a {\em simple} set. Although Post's set is bounded truth-table incomplete (more strongly, disjunctively incomplete), his question had not been settled until Friedberg \cite{Fri57} and Muchnik \cite{Muc56} finally constructed a non-recursive  Turing-incomplete {r.e.\!\!} set. Resource-bounded immune and simple sets were explicitly discussed by Flajolet and Steyaert \cite{FS74} in a general framework under the terms {\lq\lq}$\CC$-immune sets{\rq\rq} and {\lq\lq}$\CC$-simple sets{\rq\rq} for any resource-bounded complexity class $\CC$. Briefly speaking, a $\CC$-immune set is an infinite set that has no infinite subset in $\CC$ whereas a $\CC$-simple set is a set in $\CC$ whose complement is $\CC$-immune. The initial research on resource-bounded immunity and simplicity was deeply rooted in recursion theory. 

Recently, there has been a surge of renewed interests in resource-bounded immunity and simplicity and we have made a significant progress in understanding the structure of $\np$ through these notions. In this paper, taking a conventional approach toward the structural properties of the polynomial hierarchy, we extensively study the notions of resource-bounded immunity and simplicity. The purpose of this paper is to give a comprehensive guidance toward our understanding of these notions developed in computational complexity theory. We wish to present a broad spectrum of consequences obtained in the course of our study on resource-bounded immunity and simplicity. Of all complexity classes, we focus our study only on the classes lying within the polynomial hierarchy. Other classes, such as $\cequalp$ and $\dexp$, have been studied by, \eg Rothe \cite{Rot98} and Schaefer and Fenner \cite{SF98}. The goals of our investigation are to (i) analyze the behaviors of variants of immunity and simplicity notions, (ii) study the relationships between immunity and other complexity-theoretical notions, and (iii) explore new directions for better understandings of the polynomial hierarchy.  

In an early work of Ko and Moore \cite{KM81}, a number of structural properties of $\p$-immune sets were obtained; for instance, a $\p$-immune set exists even in the complexity class $\e$. Since then, there have been known close connections between immune sets and various other notions, such as complexity core of Lynch \cite{Lyn75} and instance complexity of Orponen, Ko, Sch{\"o}ning, and Watanabe \cite{OKSW94}. An additional notion of {\em $\p$-bi-immunity} was considered by Balc{\'a}zar and Sch{\"o}ning \cite{BS85}, following the previous notions of {\em almost everywhere} complexity and {\em polynomial approximation} algorithms. Departing from recursion theory, structural complexity theory has developed its own immunity notions using resource-bounded computations. Balc{\'a}zar and Sch{\"o}ning \cite{BS85} introduced the notion of {\em strongly
$\p$-bi-immune} sets and showed the existence of such sets within
$\e$. The notion of {\em almost $\p$-immune} sets was introduced by Orponen \cite{Orp86} and Orponen, Russo, and Sch{\"o}ning \cite{ORS86}. The complementary set of any almost $\p$-immune set is  called a $\p$-levelable set by Orponen, \etalc~\cite{ORS86} (where the term ``levelable'' was suggested by Ko \cite{Ko85}). Recently, Yamakami \cite{Yam95} and Schaefer and Fenner \cite{SF98} studied resource-bounded hyperimmune sets. 

Unfortunately, the existence of an $\np$-simple set is unknown because the unproven consequence $\np\neq\co\np$ immediately follows. We introduce the weaker notion of almost $\np$-simple sets and show that an $\np$-simple set exists if and only if an almost $\np$-simple set exists in $\p$. This notion also has a connection to Uspenskii's \cite{Usp57} notion of pseudosimple sets. By contrast, relativized results on the existence of simple sets abound. Homer and Maass \cite{HM83} and Balc{\'a}zar \cite{Bal85} exhibited relativized worlds where $\np$-simple sets actually exist. Homer and Maass also showed that the statement $\p\neq\np$ is not sufficient for the existence of an $\np$-simple set. The recursive oracle result of Balc{\'a}zar was expanded by Vereshchagin \cite{Ver93} to the existence of $\np$-simple sets relative to a random oracle. For each level $k$ of the polynomial hierarchy, Bruschi \cite{Bru92} constructed various relativized worlds where, for example, a $\sigmah{k}$-simple set exists. We further prove that, at each level $k$ of the polynomial  hierarchy, a strongly $\sigmah{k}$-simple set exists relative to a certain recursive oracle. In addition, we show that, relative to a generic oracle, a strongly $\np$-simple set exists. This immediately implies that an $\np$-simple set exists relative to a generic oracle. Recently, Schaefer and Fenner \cite{SF98} showed that even honestly $\np$-hypersimple sets exist relative to a generic oracle. In addition, we show the existence of a honestly $\np$-hypersimple set relative to a random oracle. We also construct a relativized world, for each $k$, in which an honest $\sigmah{k}$-hypersimple set exists.

The relationship between the simplicity notions and the reducibilities has been a central focal point in the recent literature. We can view reducibility as an algorithmic way to encode the essential information of a given set into another set. It is known that, under an appropriate assumption, strong reducibility such as Turing reducibility makes $\np$-simple sets {\lq\lq}incomplete{\rq\rq} for $\np$; that is, simple sets cannot be complete under such reductions. With regard to non-honest reductions, Hartmanis, Li, and Yesha \cite{HLY86} first showed that if $\np\cap\co\np\nsubseteq \bigcap_{\epsilon>0}\dtime{2^{n^{\epsilon}}}$ then there is no $\p$-m-complete $\np$-simple set for $\np$. Recently, Schaefer and Fenner \cite{SF98} demonstrated that no $\np$-simple set can be $\p$-1tt-complete for $\np$ unless $\up\subseteq
\bigcap_{\epsilon>0}\dtime{2^{n^{\epsilon}}}$, where $\up$ is Valiant's \cite{Val76} unambiguous polynomial-time complexity class. In this paper, we improve these results by showing that no $\sigmah{k}$-simple set can be $\deltah{j}$-1tt-complete unless $\unique(\sigmah{k}\cap\pih{k})\subseteq \subdeltaexph{\max\{j,k\}}$ for any positive integers $i,j,k$, where $\unique(\sigmah{k}\cap\pih{k})$ is the unambiguous version of $\sigmah{k}\cap\pih{k}$ introduced by Yamakami \cite{Yam92} and $\subdeltaexph{i}$ is the $i$th $\Delta$-level of the subexponential (time) hierarchy. Moreover, we show without any unproven assumption that no $\sigmah{k}$-hyperimmune set can be $\p$-T-complete for $\sigmah{k}$.

Since the existence of an $\np$-simple set is not known, Homer \cite{Hom86} looked into much weaker notions of $\np$-immunity and $\np$-simplicity by defining {\em $k$-immune} sets and {\em $k$-simple} sets within $\np$ using $O(n^k)$-time bounded nondeterministic computations. He demonstrated the existence of a $k$-simple set for each positive integer $k$. We further demonstrate the existence of a $k$-simple set that is not feasibly $k$-simple, where a {\em feasibly $k$-simple} set is an analogue of an effectively simple set in recursion theory. Employing a diagonalization argument, we can construct a feasibly $k$-immune set in $\deltah{2}$. In connection to Homer's result, we propose a working hypothesis, the so-called {\em k-immune hypothesis}: every infinite $\np$ set has an infinite subset recognized by $O(n^k)$-time nondeterministic Turing machines for a certain integer $k\geq1$. The existence of such a number $k$ yields the existence of $\np$-simple sets. This hypothesis may propel the study of $\np$-simple sets. We then show that, relative to a generic oracle, the k-immune hypothesis fails.

\paragraph{Organization of the Paper.}
The rest of this paper is organized as follows: we begin with the $\CC$-immunity and $\CC$-simplicity notions in Section \ref{sec:immune-simple} and then formulate the notions of strong $\CC$-immunity and strong $\CC$-simplicity in Section \ref{sec:strong-immune}, almost $\CC$-immunity and almost
$\CC$-simplicity in Section \ref{sec:almost-immune}, and
$\CC$-hyperimmunity and $\CC$-hypersimplicity in Section
\ref{sec:hyperimmune}. Moreover, the $k$-immunity and $k$-simplicity and their variants are discussed in Section \ref{sec:k-immune}. In the same section, the k-immune hypothesis is studied. Finally, we discuss the relationships between non-honest completeness notions and simplicity in Section \ref{sec:complete}. 

\section{Basic Notions and Notation}\label{sec:notation}

Throughout this paper, we use standard notions and notation found in the most introductory textbooks of recursion theory (\eg Odifreddi \cite{Odi89}) and computational complexity theory (\eg Du and Ko \cite{DK00}). For simplicity, we set our alphabet $\Sigma$ to be $\{0,1\}$ since this restriction does not affect the results of this paper. The reader who is already familiar with computational complexity theory and recursion theory may skip the most of this basic section.

\paragraph{Numbers and Strings.}
Let $\integer$ be the set of all integers and let $\nat$ (or $\omega$) be the set of all natural numbers (\ie nonnegative integers). Let
$\nat^{+}=\nat\setminus \{0\}$. For any finite set $A$, the notation $|A|$ denotes the {\em cardinality} of $A$. For any two integers $m,n$ with $m\leq n$, the {\em integer interval} $\{m,m+1,m+2,\ldots,n\}$ is denoted $[m,n]_{\integer}$. All {\em logarithms} are taken to base 2 and a {\em polynomial} means a univariate polynomial with integer coefficients. For convenience, we set $\log{0}=0$.  The {\em tower of $2$s} is defined as follows. Let $2_0=1$ and let $2_{n+1}=2^{2_{n}}$ for any integer $n\in\nat$. We set $Tower=\{2_n\mid n\in\nat\}$. Later, we will introduce its variant $\hat{T}$.

As noted before, we set our alphabet $\Sigma$ to be $\{0,1\}$ throughout this paper unless otherwise stated. A {\em string over $\Sigma$} is a finite sequence of symbols drawn from $\Sigma$. In particular, the {\em empty string} is denoted $\lambda$. The notation $|s|$ represents the {\em length} of a string $s$; that is, the number of symbols in $s$. The notation $\Sigma^*$ denotes the collection of all strings over $\Sigma$. Similarly, for any fixed single symbol $a$, $\{ a \}^{\ast}$ denotes the set $\{a^{i}\mid i\in\omega\}$, where $a^i$ is a shorthand for the $i$ repetitions of $a$. Moreover, for any subset $A$ of $\Sigma^*$, the notation $aA$ denotes the set $\{ax\mid
x\in A\}$.  We often identify a nonnegative integer $n$ with the
$(n+1)$th string in the standard lexicographic order on $\Sigma^{*}$: $\lambda<0<1<00<01<10<\cdots$ (sorted first by length and then lexicographically). For example, $0$ denotes $\lambda$, $3$ is $00$, and $7$ is $000$. In this order, for any string $x$, $x^{-}$ ($x^{+}$, resp.) represents the {\em predecessor} ({\em successor}, resp.) of $x$ if one exists. Conventionally, we identify a set $A$ with its {\em characteristic function}, which is defined by $A(x)=1$ if $x\in A$,  and $A(x)=0$ otherwise.

A subset of $\Sigma^*$ is called a {\em language} or simply a {\em set}. For such a set $A$, $\Sigma^*\setminus A$ is the {\em complement} of $A$ and is denoted $\overline{A}$. For each number $n\in\nat$, we write $\Sigma^{n}$,
$\Sigma^{\leq n}$, and $\Sigma^{<n}$ to denote the collections of all strings of length $n$, length $\leq n$, and length $<n$, respectively. For any sets $A$ and $B$, the notation $A\oplus B$ stands for the set $ \{ u0 \mid u \in A \} \cup \{ v1 \mid v \in B \} $, the {\em disjoint union} (or {\em join}) of $A$ and $B$.  Let $A\triangle B=(A\setminus B)\cup(B\setminus A)$, and the notation $A=^{\ast}B$ means $A\triangle B$ is finite.  We say that a complexity class $\CC$ is {\em closed under finite variations} if, for all sets $A$ and $B$, $A=^{*}B$ and $A\in\CC$ imply $B\in\CC$. The {\em census function} of $A$, denoted $cens_{A}$, is the function defined by $cens_{A}(n)=|A\cap\Sigma^{\leq n}|$ for all $n\in\nat$. A set $S$ is {\em polynomially sparse} ({\em sparse}, for short) if there exists a polynomial $p$ such that $cens_{A}(n)\leq p(n)$ for all $n\in\nat$. A {\em tally} set is a subset of either $\{0\}^*$ or $\{1\}^*$.

\paragraph{Turing Machines, Complexity Classes, INDEX$_{(k)}$, and NP$_{(k)}$.}
In this paper, we use a standard {\em multi-tape off-line Turing machine} ({\em TM}, in short) as a mathematical model of computation throughout this paper. A TM is used as an {\em acceptor} (which recognizes a language) or a {\em transducer} (which computes a function). We assume the reader's familiarity with various types of TMs (see the textbooks of Balc{\'a}zar \etalc~\cite{BDG88} and of Du and Ko \cite{DK00} for their definitions and properties). Of particular interest are deterministic, nondeterministic, and alternating TMs.  In particular, we say that an alternating TM has {\em $k$ alternations} if every computation path on every input has at most $k$ alternating states of $\forall$ and $\exists$. A {\em $\Sigma_{k}$-machine} refers to an alternating TM with $k$ alternations starting in an $\exists$ state. In particular, a $\Sigma_1$-machine is a nondeterministic TM. For convenience, we define a $\Sigma_0$-machine to be just a deterministic TM. Similarly, a $\Delta_{k}$-machine is a deterministic TM with access to an oracle which is recognized by a certain $O(n)$-time bounded $\Sigma_{k-1}$-machine. Moreover, we use an {\em unambiguous TM}, which always has at most one accepting path. For convenience, let $\CC$ be a complexity class defined by a certain type of a TM. Generally, we will use the term {\lq\lq}$\CC$-machine{\rq\rq} to mean a TM that is used to define a set in $\CC$. For instance, an {\lq\lq}$\np$-machine{\rq\rq} refers to a polynomial-time nondeterministic TM, where $\np$ is the class of all languages that can be recognized by polynomial-time nondeterministic TMs. Conventionally, we identify the acceptance and rejection of a TM with 1 and 0, respectively. Hence, for any acceptor $M$ and any input $x$, we often write $M(x)=0$ ($M(x)=1$, resp.) to mean that $M$ accepts (rejects, resp.) input $x$. 

For any TM $M$ and any function $t$ from $\nat$ to $\nat$, the notation $M(x)_{t}$ denotes the outcome of $M(x)$ if $M$ halts on input $x$ in at most $t(|x|)$ steps and outputs either $0$ (rejection) or $1$ (acceptance); otherwise, $M(x)_t$ is undefined. For any set $A$, we use the notation $M(x)\simeq A(x)$ to mean that either $M$ halts on input $x$ and outputs $A(x)$ or $M$ does not halt on input $x$. We say that $M$ is {\em consistent with $A$} or shortly {\em $A$-consistent} (denoted $M\simeq A$) if $M(x)\simeq A(x)$ for all but finitely many strings $x$. Moreover, $M$ is {\em $A$-consistent within time $t(n)$} (denoted $M\simeq_{t}A$) if $M(x)_t\simeq A(x)$ for all but finitely many strings $x$.

For any oracle TM $M$, any set $A$, and any string $x$, the notation
$Q(M,A,x)$ denotes the set of all words queried by $M$ on input $x$ to oracle $A$ and $L(M,A)$ denotes the language recognized by $M$ with oracle $A$. The notation $\mathrm{DTIME}^A(t(n))$ denotes the collection of all languages $L(M,A)$ for a constant $c>0$ and a certain deterministic oracle TM $M$ running within time $ct(n)+c$, where $n$ is the length of inputs. Similarly, we define $\mathrm{NTIME}^A(t(n))$ using nondeterministic TMs. In particular, when $A=\setempty$, we omit the superscript $A$ and simply write $\mathrm{DTIME}(t(n))$ and $\mathrm{NTIME}(t(n))$.

This paper studies only complexity classes whose computational resources are limited to polynomial time or exponential time. The basic complexity classes of our interests are given as follows. For an arbitrary oracle $A$, let $\p^A$ and $\np^A$ be respectively $\bigcup_{k\in\nat}{\rm DTIME}^A(n^k)$ and $\bigcup_{k\in\nat}{\rm NTIME}^A(n^k)$. Similarly, let $\e^A$ and $\nde^A$ be respectively $\bigcup_{c\in\nat}{\rm DTIME}^A(2^{cn})$ and $\bigcup_{c\in\nat}{\rm
NTIME}^A(2^{cn})$. Furthermore, $\dexp^A$ and $\subexp^A$ denote  $\bigcup_{k\in\nat}{\rm DTIME}^A(2^{n^{k}})$ and $\bigcap_{\epsilon>0} \mathrm{DTIME}^A (2^{n^{\epsilon}})$, respectively. As noted before, when $A=\setempty$, we omit superscript $A$. Valiant's \cite{Val76} class $\up$ is defined by polynomial-time unambiguous TMs instead of nondeterministic TMs. Later, we will introduce a more general unambiguous complexity class $\unique(\CC)$. The {\em relativized polynomial hierarchy} relative to oracle $A$ consists of the following complexity classes: $\deltah{0}(A)=\sigmah{0}(A)=\pih{0}(A)=\p^A$,
$\sigmah{k+1}(A)=\np^{\sigmah{k}(A)}$, and $\pih{k+1}(A)=\co\sigmah{k+1}(A)$ for each $k\in\nat$. The notation
$\ph^A$ denotes $\bigcup_{k\in\nat}(\sigmah{k}(A)\cup\pih{k}(A))$. If $A=\setempty$, then we obtain the (unrelativized) polynomial  hierarchy $\{\deltah{k},\sigmah{k},\pih{k}\mid k\in\nat\}$ by dropping superscript $A$. In this paper, we define only the $\Delta$-levels of the {\em exponential (time) hierarchy} and the {\em subexponential (time) hierarchy}: $\deltaexph{0}=\dexp$, $\deltaexph{k+1} = \dexp^{\sigmah{k}}$, $\subdeltaexph{0}= \subexp$, and $\subdeltaexph{k+1} = \subexp^{\sigmah{k}}$ for every $k\in\nat$.

We assume a standard effective enumeration $\{\varphi_{i}\}_{i\in\omega}$ of all nondeterministic TMs (with repetitions). Each index $i$ of such a machine $\varphi_i$ is conventionally called the {\em G{\"o}del number} of the machine $\varphi_i$. For each index $i$, we define the set $W_{i}=\{x \mid \varphi_{i}(x)\!\!\downarrow=1\}$, where {\lq\lq}$\varphi_{i}(x)\!\!\downarrow${\rq\rq} means that $\varphi_{i}$ eventually halts on input $x$.  Fix $k\in\nat$. Let $\np_{(k)}$ be the collection of all sets $W_{i}$ such that, for any string $x\in W_i$, the running time of $\varphi_{i}$ on input $x$ is at most $|i|\cdot|x|^{k}+|i|$, where the notation $|i|$ means the length of the string that is identified with $i$ and thus  $|i|=\floors{\log(i+1)}$. 
Note that $\np=\bigcup_{k\in\nat}\np_{(k)}$. We set
$\gindex_{(k)}=\{i\in\omega \mid W_{i}\in\np_{(k)}\}$. For any string $x\in\Sigma^*$ and any set $A\subseteq\Sigma^*$, let $C^A(x)$ denote the {\em relativized Kolmogorov complexity of $x$ relative to $A$}; that is, the minimal length of indices $e$ such that $M^A_e(\lambda)=x$, where $M^A_e$ is the $e$th deterministic TM (without any specific time bound). 

We often express a finite sequence of strings by a single string. Let $\Sigma^{<\omega}$ denote the set of all finite sequences of strings. We assume a bijection $\pair{\hs{2}}$ from $\Sigma^{<\omega}$ to $\Sigma^*$ that is polynomial-time computable and polynomial-time invertible. For such an encoded sequence $s=\pair{x_1,x_2,\ldots,x_k}$, $set(s)$ denotes the corresponding set $\{x_1,x_2,\ldots,x_k\}$.

\paragraph{Partial Functions, $\fdeltah{k}$, $\sigmah{k}\mathrm{SV}$, and $\unique(\sigmah{k}\cap\pih{k})$.}
In this paper, we mainly use ``partial'' functions and all functions are assumed to be single-valued although, in other literature, more general ``multi-valued'' functions are discussed. It is important to note that total functions are also partial functions. For any partial function $f$, the notations $\dom(f)$ and $\ran(f)$ denote the {\em domain} of $f$ and the {\em range} of $f$, respectively. Let $f$ be any partial function from $\Sigma^*$ to $\Sigma^*$. We say that $f$ is {\em lexicographically increasing} if, for any string $x\in\dom(f)$, $f(x)$ is greater than $x$ in the standard lexicographical order on $\Sigma^*$.  Moreover, $f$ is {\em polynomially bounded} if there exists a polynomial $p$ such that $|f(x)| \leq p(|x|)$ for any $x \in \dom (f)$. By contrast, $f$ is called {\em polynomially honest} ({\em honest}, in short) if there exists a polynomial $p$ such that $|x|\leq p(|f(x)|)$ for any $x\in\dom(f)$. Whenever it is clear from the context that $f(x)$ is of the form $\pair{y_1, \ldots, y_k}$ for every $x\in\dom(f)$ (where $k$ may depend on $x$), we say that $f$ is {\em componentwise honest} if there exists a polynomial $p$ such that $|x| \leq p(|y_j|)$ for any string $x\in\dom(f)$ and any number $j\in [1,k]_{\integer}$ with $k\geq1$.

To avoid notational mess, we define standard function classes as
collections of {\em partial} functions and, whenever we need total
functions, we explicitly indicate the {\em totality} of functions but use the same notation. For instance, $\fp$ denotes the set of all single-valued {\lq\lq}partial{\rq\rq} functions computable deterministically in polynomial time. Now, fix $k\in\nat^{+}$. More generally, we use the notation $\fdeltah{k}$ for the collection
of all single-valued partial functions $f$ such that there exist a set $B\in\sigmah{k-1}$ and a polynomial-time deterministic oracle TM $M$ satisfying the following condition: for every $x$, if
$x\in\dom(f)$ then $M^B(x)$ halts with an accepting state and outputs
$f(x)$ and, otherwise, $M^B(x)$ never halts in any accepting state (in this case, $f(x)$ is {\em undefined}). For technical convenience, we may assume that such an oracle TM $M$ is {\em clocked} in an appropriate manner; therefore, we can force $M^B(x)$ to halt in a rejecting state whenever $f(x)$ is undefined. Clearly, $\fdeltah{1}$ coincides with $\fp$. A single-valued partial function $f$ is in $\npsv$ if there is an $\np$-machine $M$ such that if $x\not\in \dom(f)$ then all the computation paths of $M(x)$ terminate with   rejecting configurations, and otherwise, $M(x)$ terminates with at least one accepting configuration and $M(x)$ outputs $f(x)$ along {\em all} accepting computation paths. The suffix {\lq\lq}SV{\rq\rq} in $\sigmah{k}\mathrm{SV}$ stands for {\lq\lq}single-valued.{\rq\rq} We expand $\npsv$ to $\sigmah{k}\mathrm{SV}$ in the following machine-independent way. First, set $\sigmah{0}\mathrm{SV}=\fp$ for convenience. For any partial function $f$ from $\Sigma^*$ to $\Sigma^*$, the {\em graph} of $f$ is the set $Graph(f)=\{\pair{x,f(x)}\mid x\in\dom(f)\}$. For each number $k\in\nat^{+}$, let $\sigmah{k}\mathrm{SV}$ denote the class of all single-valued partial functions $f$ such that $f$ is polynomially bounded and $Graph(f)$ is in $\sigmah{k}$. It is not difficult to prove that $\sigmah{1}\mathrm{SV}$ indeed coincides with $\npsv$. To emphasize the total functions, we use the notation $\sigmah{k}\mathrm{SV}_t$ to denote the collection of all {\em total} functions in $\sigmah{k}\mathrm{SV}$. The reader may refer to Selman's \cite{Sel94} work on relationships among the aforementioned function classes.

Using the graphs of partial functions, Yamakami \cite{Yam92} defined the class operator $\mathrm{U}$ as follows. For any complexity class $\CC$ of languages, $\unique(\CC)$ (or simply $\unique\CC$) is the collection of all sets $A$ such that there exists a single-valued partial function $f$ satisfying the following three conditions: (i) $f$ is polynomially bounded, (ii) $A=\dom(f)$, and (iii)  $Graph(f)\in\CC$. In particular, we obtain $\unique\deltah{k}$ and $\unique(\sigmah{k}\cap\pih{k})$ for each $k\in\nat$. Notice that $\unique\deltah{1}$ coincides with $\up$. Fortnow and Yamakami \cite{FY96} showed the generic separation between $\unique\deltah{k}\cap\pih{k}$ and $\deltah{k}$ for every $k\geq2$.

\paragraph{Reductions and Quasireductions.}
Fix $k\in\nat$ and let $A$ and $B$ be any subsets of $\Sigma^*$. A partial function $f$ from $\Sigma^*$ to $\Sigma^*$ is called a {\em $\sigmah{k}$-m-quasireduction} ({\em $\deltah{k}$-m-quasireduction}, resp.) {\em from $A$ to $B$} if (i) $f$ is in $\sigmah{k}\mathrm{SV}$ ($\fdeltah{k}$, resp.), (ii) $\dom (f)$ is infinite, and (iii) for any string $x \in \dom(f)$, $x \in A$ iff $f(x) \in B$. If in addition $f$ is {\em total}, then $f$ is called a {\em $\sigmah{k}$-m-reduction} ({\em $\deltah{k}$-m-reduction}, resp.) {\em from $A$ to $B$}, and we say that $A$ is $\sigmah{k}$-m-reducible ($\deltah{k}$-m-reducible, resp.) to $B$. If $A$ is $\deltah{k}$-m-reducible to $B$ via a one-to-one reduction $f$, we say that $A$ is {\em $\deltah{k}$-1-reducible} to $B$ via $f$. In contrast, $A$ is {\em $\deltah{k}$-T-reducible} to $B$ if there exists an oracle $\deltah{k}$-machine $M$ which recognizes $A$ with access to $B$ as an oracle. Moreover, we define bounded reducibilities as follows. For notational convenience, for any set $B$ and any function $f$ from $\Sigma^*$ to $\Sigma^*$, we abbreviate as $B(f(x))$ the $k$-bit string $B(y_1)B(y_2)\cdots B(y_k)$ if $f(x) = \pair{y_1,y_2, \ldots, y_k}$, where $k$ is called the {\em norm} of $f$ at $x$.
A set $A$ is called {\em $\deltah{k}$-tt-reducible} to $B$ via $(\nu,f,\alpha)$ if (i) $\nu$ is a total $\fdeltah{k}$-function from $\Sigma^*$ to $\{0\}^*$, (ii) $f$ is a total $\fdeltah{k}$-function from $\Sigma^{\ast}$ to $\Sigma^{\ast}$ such that $|\nu(x)|$ is the norm of $f$ at every $x\in\Sigma^*$, and (iii) $\alpha$ is a total $\fdeltah{k}$-function from $\Sigma^{\ast} \times \Sigma^{\ast}$ to $\{ 0,1 \}$ such that, for every $x\in\Sigma^*$, $x\in A$ iff $\alpha (x,B(f(x))) = 1$. For any constant $i\in\nat^{+}$, $A$ is {\em $\deltah{k}$-$i$tt-reducible} to $B$ via $(f,\alpha)$ if $A$ is $\deltah{k}$-tt-reducible to $B$ via $(\nu,f,\alpha)$, where $\nu (x) = 0^i$ for all $x$; in other words, the norm of this reduction is always $i$ at any $x$. A set $A$ is {\em $\deltah{k}$-btt-reducible} to $B$ if there exists a number $i\in\nat^{+}$ such that $A$ is $\deltah{k}$-$i$tt-reducible to $B$.  A set $A$ is {\em $\deltah{k}$-d-reducible to $B$} via $f$ if (i) $f$ is a total $\fdeltah{k}$-function and (ii) for every $x\in\Sigma^*$, $x \in A$ iff $B \cap set(f(x)) \neq \setempty$. In contrast, $A$ is {\em
$\deltah{k}$-c-reducible to $B$} via $f$ if $\overline{A}$ is $\deltah{k}$-d-reducible to $\overline{B}$ via $f$.

Next, we define honest reductions. For any $r\in\{\mathrm{m,T}\}$ ($r\in\{\mathrm{c,d}\}$, resp.), $A$ is {\em h-$\deltah{k}$-$r$-reducible to $B$}  via a reduction $f$ if $A$ is $\deltah{k}$-$r$-reducible to $B$ via $f$ such that $f$ is honest (componentwise honest, resp.). If $r\in\{\mathrm{btt,tt}\}$, then $A$ is said to be {\em h-$\deltah{k}$-$r$-reducible to $B$} via a reduction triplet $(\nu,f,\alpha)$ if $A$ is $\deltah{k}$-$r$-reducible to $B$ via $(\nu,f,\alpha)$ with an extra condition that $f$ is componentwise honest. For Turing reductions, we say that $A$ is {\em h-$\deltah{k}$-T-reducible} to $B$ via a reduction machine $M$ if $A$ is $\deltah{k}$-T-reducible to $B$ via $M$ such that $M$ makes only {\lq\lq}honest{\rq\rq} queries; more precisely, (i) $M$ runs in polynomial time, (ii) for every $x\in\Sigma^*$, $x \in A$ iff $M^B(x)=1$, and (iii) there exists a polynomial $p$ such that $p(|w|) \geq |x|$ for every $x$ and every $w \in Q(M,B,x)$. In this case, we say that the reduction machine $M$ is {\em honest}.

When $\CC\in\{\deltah{k}\mid k\in\nat^{+}\}$ and $r\in\{\mathrm{1,m,c,d,}k\mathrm{tt,btt,tt,T}\}$, notationally we write $A\leq^{\CC}_{r}B$ ($A\leq^{\mathrm{h}\mbox{-}\CC}_{r}B$, resp.) to mean that $A$ is $\CC$-$r$-reducible (h-$\CC$-$r$-reducible, resp.) to $B$. A set $S$ is called {\em $\CC$-r-hard for $\DD$} if every set in $\DD$ is $\CC$-$r$-reducible to $S$. Moreover, a set $S$ is called {\em $\CC$-r-complete for $\DD$} if $S$ is in $\DD$ and $S$ is $\CC$-$r$-hard for $\DD$. For the honest reductions, the notions of {\em h-$\CC$-r-hardness} and {\em h-$\CC$-r-completeness} are defined in a similar fashion.

Finally, we say that a complexity class $\CC$ is {\em closed downward under a reduction $\leq_r$ on infinite sets} if, for any pair of infinite sets $A$ and $B$, $A\leq_r B$ and $B\in\CC$ imply $A\in \CC$.

\paragraph{Forcing, Generic Sets, and Random Oracles.}
A {\em forcing condition} $\sigma$ is a partial function from
$\Sigma^*$ to $\{0,1\}$ such that $\dom(\sigma)$ is finite. Such a function $\sigma$ is identified with the string
$\pair{\pair{v_0,v_1,\ldots,v_{n-1}},
\linebreak
\pair{w_0,w_1,\ldots,w_{n-1}}}$, where $\dom(\sigma)=\{v_0,v_1,\ldots,v_{n-1}\}$ with $v_0<v_1<\cdots<v_{n-1}$ and $\sigma(v_i)=w_i$ for
all $i<n$. By this identification, a set of forcing conditions can be
treated as a set of strings. For two forcing conditions $\sigma$ and
$\tau$, we say that $\tau$ {\em extends} $\sigma$ (denoted $\sigma\subseteq\tau$) if $\dom(\sigma)\subseteq\dom(\tau)$ and $\sigma(x)=\tau(x)$ for all
$x\in\dom(\sigma)$. Similarly, for any subset $A$ of $\Sigma^*$, we say that $A$ {\em extends} $\sigma$ (denoted $\sigma\subseteq A$) if $\sigma(x)=A(x)$, viewed as the characteristic function, for all $x\in\dom(\sigma)$. A set $S$ of forcing conditions is {\em dense along $A$} if every forcing condition $\sigma\subseteq A$ has an extension $\tau$ in $S$. In particular, we say that $S$ is {\em dense} if $S$ is dense along every subset $A$ of $\Sigma^*$. A set $L$ {\em meets} a set $S$ of forcing conditions if $L$ extends a certain forcing condition in $S$. For any complexity class $\CC$, a set $L$ is called {\em $\CC$-generic} if $L$ meets every set in $\CC$ that is dense along $L$. In particular, when $\CC$ is the class of all arithmetical sets, we use the conventional term {\lq\lq}Cohen-Feferman generic{\rq\rq} instead. When a generic set is specifically used as an oracle, it is customary called a {\em generic oracle}.

For any set $A\subseteq\{0,1\}^*$, we identify $A$ with its characteristic sequence $\alpha_A=A(\lambda)A(0)A(1)A(00)\cdots$; namely, the $i$th bit of $\alpha_A$ is the value of $A$ on the lexicographically $i$th string on $\Sigma^*$. Such an infinite sequence corresponds to a real number in the unit interval $[0,1]$. Let $r_{A}$ be the real number whose binary expansion is of the form $0.\alpha_A$. Notice that some real numbers have two equivalent binary expansions (\eg $0.01$ and $0.00\dot{1}$, where $\dot{1}=11\cdots1\cdots$); however, because the set $\{0.s\dot{1}\mid s\in\Sigma^*\}$ has the Lebesgue measure $0$ in $[0,1]$, we can ignore this duality problem for our purpose. In general, let $\varphi(X)$ be any mathematical property with a variable $X$ running over all subsets of $\{0,1\}^*$. We say that {\em $\varphi(X)$ holds (with probability $1$) relative to a random oracle $X$} if the set $\{r_{A}\mid A\subseteq\{0,1\}^*\:\wedge\: \varphi(A)\mbox{ holds}\}$ has Lebesgue measure 1 in the interval $[0,1]$.

\section{Immunity and Simplicity}
\label{sec:immune-simple}

The original notions of immunity and simplicity date back to the mid
1940s. Post \cite{Pos44} first constructed a simple set for
the class $\mathrm{RE}$ of all {r.e.\!\!} sets.  The new breed of resource-bounded immunity and simplicity waited to be introduced until mid
1970s by an early work of Flajolet and Steyaert \cite{FS74}. In their seminal paper, Flajolet and Steyaert constructed various recursive sets that, for instance, have no infinite $\mathrm{DTIME}(t(n))$-subsets, and they introduced the term {\lq\lq}$\mathrm{DTIME}(t(n))$-immune sets{\rq\rq} for such sets. Later, Ko and Moore \cite{KM81} studied the polynomial-time bounded immunity, which is now called $\p$-immunity. Subsequently, Balc{\'a}zar and Sch{\"o}ning \cite{BS85}, motivated by Berman's \cite{Ber77} work, considered $\p$-bi-immune sets, which are $\p$-immune sets whose complements are also $\p$-immune. Homer and Maass \cite{HM83} extensively discussed the cousin of $\p$-immune sets, known as $\np$-simple sets. The importance of these notions was widely recognized through the 1980s. Since these notions can be easily expanded to any complexity class $\CC$, we begin with an introduction of the general notions of $\CC$-immune sets, $\CC$-bi-immune sets, and $\CC$-simple sets. They are further expanded in various manners in later sections.

\begin{definition} 
Let $\CC$ be any complexity class.
\vs{-2}
\begin{enumerate}
\item A set $S$ is {\em $\CC$-immune} if $S$ is infinite and
there is no infinite subset of $S$ in $\CC$. 
\vs{-2}
\item A set $S$ is {\em $\CC$-bi-immune} if $S$ and $\overline{S}$
are both $\CC$-immune. 
\vs{-2}
\item A set $S$ is {\em $\CC$-simple} if $S$ belongs to $\CC$ and $\overline{S}$ is
$\CC$-immune.
\end{enumerate}
\end{definition}

We sometimes use the term {\lq\lq}$\CC$-coimmune{\rq\rq} to mean that the complement of a given set is $\CC$-immune. Using this term, a $\CC$-simple set is a $\CC$-coimmune $\CC$-set. Clearly, the intersection of any two $\CC$-immune sets is either $\CC$-immune or finite. Note that the existence of a $\CC$-simple set immediately implies $\CC\neq\co\CC$; however, the separation $\CC\neq\co\CC$ does not necessarily guarantee the existence of $\CC$-simple sets. In this paper, we focus our study only on the complexity classes lying in the polynomial hierarchy.

It is well-known that $\p$-immune sets exist even in the class $\e$. In particular, Ko and Moore \cite{KM81} constructed a $\p$-immune set that is also $\p$-tt-complete for $\e$.  Note that no h-$\p$-m-complete set for $\np$ can be $\p$-immune since the inverse image of a $\p$-immune set by any h-$\p$-m-reduction is either $\p$-immune or finite. Using a relativization technique, Bennett and Gill \cite{BG81} proved that a $\p$-immune set exists in $\np$ relative to a random oracle. A recursive oracle relative to which $\np$ contains $\p$-immune sets was constructed later by Homer and Maass \cite{HM83}. Torenvliet and van Emde Boas \cite{TB89} strengthened their result by demonstrating a relativized world where $\np$ has a $\p$-immune set which is also $\np$-simple. The difference between sparse immune sets and tally immune sets was discussed by Hemaspaandra and Jha \cite{HJ95}, who constructed an oracle relative to which $\np$ has a $\p$-immune sparse set but no $\p$-immune tally set. Taking a different approach, Blum and Impagliazzo \cite{BI87} proved the existence of $\p$-immune sets in $\np$ relative to a Cohen-Feferman generic oracle. Relative to a random oracle, Hemaspaandra and Zimand \cite{HZ96} showed that $\np$ contains even M{\"u}ller's \cite{Mul93} notion of a $\p$-balanced immune set, which contains asymptotically half of the elements of each infinite $\p$-set. As for sets in the Boolean hierarchy over $\np$, Cai \etalc~\cite{CGH+88} proved that they are neither $\np$-bi-immune, $\co\np$-bi-immune, nor $\bh$-immune, where $\bh$ is the union of all classes in the Boolean hierarchy within $\np$. 

The notion of $\CC$-immunity is related to various other notions, which include complexity cores of Lynch \cite{Lyn75} and instance complexity of Orponen, Ko, Sch{\"o}ning, and Watanabe \cite{OKSW94}. Balc{\'a}zar and Sch{\"o}ning \cite{BS85} showed that a set $S$ is $\p$-bi-immune exactly when $\Sigma^{*}$ is a complexity
core for $S$. A set $A$ is called a {\em $\sigmah{k}$-hardcore for $X$} if, for every polynomial $p$ and for every $\Sigma_{k}$-machine $M$ recognizing $A$, there exists a finite subset $S$ of $X$ such that $M(x)_{p}$ is undefined for all $x\in X \setminus S$. Similarly, we can define the notion of a $\deltah{k}$-hardcore for $X$ using  $\Delta_{k}$-machines instead of $\Sigma_{k}$-machines. Furthermore, as Orponen {et al.} proved, instance complexity also characterizes $\p$-immunity. Let $\CC$ be any
complexity class in the polynomial hierarchy. The {\em $t$-time
bounded $\CC$-instance complexity of $x$ with respect to $A$}, denoted
$\CC\mbox{-}ic^{t}(x:A)$, is defined to be the minimal length of an 
index $e$ such that the $\CC$-machine indexed $e$, which is
$A$-consistent within time $t(n)$, outputs $A(x)$ on input $x$ in time
$t(n)$. 

In the following lemma, we give a characterization of $\CC$-immunity by (a generalization of) the above notions. The lemma can be easily obtained by modifying the proofs of Balc{\'a}zar and Sch{\"o}ning \cite{BS85} and of Orponen \etalc~\cite{OKSW94}. 

\begin{lemma}\label{lemma:immune-core} 
Let $\CC\in\{\deltah{k},\sigmah{k}\mid k\in\nat\}$ and let $S$ be any
recursive subset of $\Sigma^*$. The following three statements are
equivalent.
\vs{-2}
\begin{enumerate}
\item $S$ is $\CC$-immune.
\vs{-2}
\item $S$ is a $\CC$-hardcore for $S$.
\vs{-2}
\item For any polynomial $p$ and any constant $c>0$, the set $\{x\in
S\mid \CC\mbox{-}ic^p(x:S)\leq c\}$ is finite. 
\end{enumerate}
\end{lemma}

Note that recursiveness of $S$ in Lemma \ref{lemma:immune-core} cannot be removed. For completeness, we include the proof of the lemma.

\begin{proofof}{Lemma \ref{lemma:immune-core}}
We show only the case where
$\CC=\sigmah{k}$ for any fixed integer $k\geq1$. Let $S$ be any recursive subset of $\Sigma^*$ and assume that $S$ is recognized by a deterministic TM $N$.
 
3 implies 2) Assume that there exist a $\Sigma_{k}$-machine
indexed $e$ and a polynomial $p$ such that $M_e$ recognizes $S$ and
$M_e(x)_p$ exists for infinitely many $x$ in $S$. Thus,
$\sigmah{k}\mbox{-}ic^p(x:S)\leq e$ for infinitely many $x$ in $S$.

2 implies 1) In this proof, we need the computability of $S$.  Assume
that $S$ is not $\sigmah{k}$-immune; namely, there exist a polynomial
$p$ and a $\Sigma_{k}$-machine $M$ running in time $p(n)$ that
recognizes a certain infinite subset of $S$. We define another machine
$M'$ as follows:
\begin{quote}
{\sf On input $x$, run $M$ on the same input $x$. Whenever $M$ halts
in an accepting state, then accept the input and halt. Otherwise, run
$N$ on input $x$ and output whatever $N$ does.}
\end{quote}
Clearly, $M'$ is a $\Sigma_{k}$-machine and recognizes $S$. Note that, for infinitely many $x$ in $S$, 
the running time of $M'$ on each input $x$ does
not exceed $p(|x|)$. Therefore, $S$ cannot be a $\sigmah{k}$-hardcore.

1 implies 3) Assume that there exist an infinite subset $A$ of $S$, a
polynomial $p$, and an integer $c>0$ such that
$\sigmah{k}\mbox{-}ic^p(x:S)\leq c$ for all strings $x$ in $A$. Thus, for every $x$ in $A$, there is a $\Sigma_{k}$-machine indexed $e$ with $0\leq |e|\leq c$, say $M_e$, such that $M_e$ is $S$-consistent within time $p$ and $M_e(x)=S(x)$. For each $e$, define $L_e=\{x\in A\mid M_e\simeq_{p}S \:\wedge\: M_e(x)_{p}=S(x)\}$. Note that
$A$ equals $\bigcup_{e:1\leq |e|\leq c} L_{e}$. Since $A$ is infinite, $L_e$ is also infinite for a certain index $e$ with $1\leq |e|\leq c$. Take such an index $e$ and define $L=\{x\mid M_e(x)_{p}=1\}$. Clearly, $L_e\subseteq L\subseteq S$ because $M_e\simeq_{p}S$. Since $L_e$ is infinite, $L$ is also infinite.  Moreover, $L$ belongs to $\sigmah{k}$
since $M_e$ is a $\Sigma_{k}$-machine.  Hence, $S$ cannot be
$\sigmah{k}$-immune.
\end{proofof}

Balc{\'a}zar and Sch{\"o}ning \cite{BS85} built a bridge between
$\p$-bi-immune sets and finite-to-one reductions, which led them
further to introduce the notion of strongly $\p$-bi-immune
sets. Expanding their argument, we give below a characterization 
of $\CC$-bi-immunity as well as $\CC$-immunity in terms of $\CC$-m-quasireductions. For any
partial function $f$ and any element $b$, let
$f^{-1}(b)=\{x\in\dom(f)\mid f(x)=b\}$. Note that
$f^{-1}(b)=\setempty$ if $b\not\in\ran(f)$. 

\begin{lemma}\label{lemma:finite-to-one}
Let $\CC\in\{\deltah{k},\sigmah{k}\mid k \in\nat\}$ and
$S\subseteq\Sigma^*$.
\vs{-2}
\begin{enumerate}
\item $S$ is $\CC$-immune if and only if (i) $S$ is infinite and 
(ii) for every set $B$, every $\CC$-m-quasireduction $f$ 
from $S$ to $B$, and every $u$ in $B$, 
the inverse image  $f^{-1} (u)$ is a finite set. 
\vs{-2}
\item $S$ is $\CC$-bi-immune if and only if $($i$)$ $S$ is infinite and 
$($ii$)$ for every set $B$, every $\CC$-m-quasireduction $f$ 
from $S$ to $B$, and every $u$, 
the inverse image $f^{-1}(u)$ is finite. 
\end{enumerate}
\end{lemma}

\begin{proof}
We show only the case where $\CC=\sigmah{k}$ for any number
$k\in\nat^{+}$. Since the second claim immediately follows from the first one, we hereafter give the proof of the first claim.

Assume that $S$ is not $\sigmah{k}$-immune. There exists an
infinite $\sigmah{k}$-subset $A$ of $S$. Take a fixed element $a_0$ in $A$. Define $f(x)=a_0$ if $x\in A$, and $f(x)$ is undefined otherwise.
It is obvious that $f \in \sigmah{k}\mathrm{SV}$.  Conversely,
assume that there exists a set $B$ and a $\sigmah{k}$-m-quasireduction
$f$ from $S$ to $B$ such that $f^{-1}(u)$ is infinite for a certain string $u \in B$. Clearly, $f^{-1}(u) \subseteq S$ and $f^{-1}(u) \in
\sigmah{k}$ because $Graph(f)$ is in $\sigmah{k}$ and thus the set  $\dom(f)=\{x\mid \exists y\in\Sigma^{|f(x)|}[(x,y)\in Graph(f)]\}$ is also in $\sigmah{k}$. Therefore, $S$ is not $\sigmah{k}$-immune.
\end{proof}

The characterizations of $\CC$-immunity given in Lemmas 
\ref{lemma:immune-core} and \ref{lemma:finite-to-one} demonstrate a 
significant role of the immunity in complexity theory. We will give a related notion, called strongly $\CC$-immunity, in Section \ref{sec:strong-immune}.

Whether an $\np$-simple set exists is one of the long-standing open
problems because such a set separates $\np$ from $\co\np$. Nonetheless, $\np$-simple sets are known to exist in various relativized worlds. In the early 1980s, Homer and Maass \cite{HM83} and Balc{\'a}zar \cite{Bal85} constructed relativized worlds where an $\np$-simple set exists. Later, Vereshchagin \cite{Ver93} proved that an $\np$-simple set exists relative to a random oracle. {}From Theorem \ref{theorem:strongly-simple} in Section \ref{sec:strong-immune}, for instance, it immediately follows that an $\np$-simple set exists relative to a Cohen-Feferman generic oracle. Torenvliet \cite{Tor88} built an oracle relative to which a $\sigmah{2}$-simple set exists. For a much higher level $k$ of the polynomial hierarchy, Bruschi \cite{Bru92} constructed an oracle relative to which $\sigmah{k}$-simple sets exist using the size lower bounds of certain non-uniform constant-depth circuits. In addition, sets being both simple and immune were studied by, \eg Buhrman and Torenvliet \cite{BT98} and Torenvliet and van Emde Boas \cite{TB89}.

In the rest of this section, we discuss closure properties of the class of all $\sigmah{k}$-immune sets because no such closure property has been systematically studied in the literature. We claim that this class is closed downward under h-$\deltah{k}$-c-reductions on infinite sets; however, we cannot replace this conjunctive reducibility by disjunctive reducibility. 

\begin{theorem}\label{theorem:reduction-immune}
Let $k\in\nat^{+}$. 
\vs{-2}
\begin{enumerate}
\item The class of all $\sigmah{k}$-immune sets 
is closed downward
under h-$\deltah{k}$-c-reductions on infinite sets. 
\vs{-2}
\item The class of all $\np$-immune sets is not closed under
h-$\p$-d-reductions or h-$\p$-2tt-reductions on infinite sets.
\end{enumerate}
\end{theorem}

\begin{proof}
1) Let $A$ and $B$ be any infinite sets and let $f$ be any
h-$\deltah{k}$-c-reduction $f$ from $A$ to $B$. Assume that $B$ is
$\sigmah{k}$-immune. We want to show that $A$ is also
$\sigmah{k}$-immune. Assume to the contrary that $A$ contains an
infinite $\sigmah{k}$-subset $C$. There exists a polynomial $p$ such
that $|x|\leq p(|y|)$ for all $x\in\Sigma^*$ and for all 
$y\in set(f(x))$. Note that the set
$\bigcup_{x\in C}set(f(x))$ is infinite since $f$ is componentwise honest. Let
$D=\{y\mid \exists x \in C [ |x| \leq p(|y|) \:\wedge\: y \in
set(f(x)) ]\}$. Notice that $D$ belongs to $\sigmah{k}$ since $f$ is in
$\fdeltah{k}$ and $C$ is in $\sigmah{k}$. Moreover, $D$ is an infinite
subset of $B$. This contradicts our assumption. Hence, $A$ is
$\sigmah{k}$-immune.

2) Define $A=\{ 0 \}^{*}$ and let $f$ be the function defined as
follows: $f(x)=\langle 0x, 1x \rangle$ for every string
$x$. We construct an $\np$-immune set $B$ to which $f$
h-$\p$-d-reduces $A$ together with an auxiliary set $C$. Let $\{ N_j
\}_{j\in\omega}$ be an effective enumeration of nondeterministic
oracle TMs whose running times are bounded above by polynomials
independent of the choice of oracles.

The desired sets $B=\bigcup_{m\in\nat}B_m$ and
$C=\bigcup_{m\in\nat}C_m$ are constructed by stages.

\begin{quote}
{\sf Stage $0$:} Let $B_0 =\setempty$ and $C_0 =\setempty$. 

{\sf Stage $m+1$:} At this stage, we wish to define 
$B_{m+1}$ and $C_{m+1}$. Find the minimal natural number
$j$ such that the following three conditions hold: (i) $L(N_j) \cap \Sigma^{\leq m} \subseteq B_m$, (ii) either $N_j(0^{m+1}) =1$ or
$N_j(10^{m}) =1$, and (iii) $j \not\in C_m$. For this $j$, define $C_{m+1}$ as $C_{m}\cup\{j\}$. The set $B_{m+1}$ is defined by the
following three cases.  If $N_j(0^{m+1}) = N_j(10^{m}) =1$, then 
let $B_{m+1}$ be $B_m \cup \{ 0^{m+1}
\}$. If $N_j(0^{m+1}) =1$ and $N_j(10^{m}) =0$, then let
$B_{m+1}=B_m \cup \{ 10^{m} \}$. If $N_j(0^{m+1}) =0$ and
$N_j(10^{m}) =1$, then let $B_{m+1}=B_m \cup \{ 0^{m+1} \}$. Note that $B_{m+1}\subseteq\Sigma^{\leq m+1}$.
\end{quote}\vs{-2}

\n By the above construction, clearly $B$ is 
infinite and, for every string $x$, $x$ is of the form $0^k$ for a certain natural number $k$ if and only if at least one of $0x$ and
$1x$ belongs to $B$. Hence, $f$ h-$\p$-d-reduces $A$ to $B$. Note
that this reduction is also an h-$\p$-2tt-reduction.

To conclude the proof, it suffices to show that $B$ has no infinite
$\np$-subset. To lead to a contradiction, we assume that $L(N_j)$ is
an infinite subset of $B$ for a certain index $j$. In case where $j
\in C$, $j$ must be in $C_{m+1}$ at a certain stage $m$. If we take
minimal such $m$, then $j$ is used at stage $m+1$, and hence $L(N_j) $ is not a subset of $B$, a contradiction. Therefore, $j \not\in
C$. Since $L(N_j)$ is infinite, there is a natural number $m$ with the
following property: (i) $N_j$ accepts a string of length $m+1$, and
(ii) every index $i<j$ such that $i \in C$ is entered in $C$ before
stage $m+1$. Since $L(N_j)$ is a subset of $B$, by (i), $N_j$ accepts
at least one of $0^{m+1}$ and $10^m$. By (ii), $j$ is used at stage $m+1$. This implies that $L(N_j)$ is not a subset of $B$, a contradiction. Therefore, $B$ has no infinite $\np$-subset.
\end{proof}

If $\CC$-simple sets exist for a class $\CC$, then how complex are they? Such $\CC$-simple sets have a nonempty intersection with every infinite set in $\CC$ and thus they cannot be {\lq\lq}complete{\rq\rq} under certain types of reductions. As an immediate consequence of Theorem \ref{theorem:reduction-immune}(1), we obtain the following corollary. 

\begin{corollary}\label{cor:reduction-immune}
Let $k\in\nat^{+}$. 
No $\sigmah{k}$-simple set is h-$\deltah{k}$-d-complete 
for $\sigmah{k}$.
\end{corollary}

\begin{proof}
Assume that $B$ is $\sigmah{k}$-simple and
h-$\deltah{k}$-d-complete for $\sigmah{k}$. Notice that $\overline{B}$
is $\sigmah{k}$-immune. Fix any infinite coinfinite set $A$ in $\p$. Note that
$\overline{A}$ is h-$\deltah{k}$-c-reducible to $\overline{B}$ because
of the completeness of $B$. {}From Theorem
\ref{theorem:reduction-immune}(1), it follows that $\overline{A}$ is
$\sigmah{k}$-immune. By the immunity condition, $\overline{A}$ is not
in $\sigmah{k}$, a contradiction. Thus, $B$ cannot be 
$\sigmah{k}$-simple.
\end{proof}

Recently, Agrawal (cited in \cite{SF98}) showed, using the
$\np$-levelability of $\overline{\mathrm{SAT}}$ (assuming
$\mathrm{SAT}\not\in\p$), that no $\np$-simple set is
h-$\p$-btt-complete for $\np$, where SAT is the set of all satisfiable Boolean formulas. His argument will be generalized in
Section \ref{sec:hyperimmune}.

\section{Strong Immunity and Strong Simplicity}
\label{sec:strong-immune}

Following the introduction of $\p$-bi-immunity, Balc{\'a}zar and
Sch{\"o}ning \cite{BS85} stepped forward to introduce the notion of strong $\p$-bi-immunity, which comes from the
quasireducibility-characterization of $\p$-bi-immunity given in Lemma
\ref{lemma:finite-to-one}(2). While $\p$-bi-immunity requires its 
quasireductions to be finite-to-one, strong $\p$-bi-immunity requires
the quasireductions to be almost one-to-one, where a quasireduction
$f$ is called {\em almost one-to-one} on a set $S$ if the {\em collision set} $\{ (x, y) \in (\dom (f) \cap S)^2
\mid x < y \:\wedge\: f(x)=f(y) \} $ is finite. As Balc{\'a}zar and Sch{\"o}ning demonstrated, such strongly
$\p$-bi-immune sets indeed exist in $\e$. Resource-bounded genericity also implies strong immunity
as shown in Proposition \ref{prop:generic-strong}.

Generalizing the notion of $\p$-bi-immunity, we can introduce strong $\CC$-bi-immunity for any complexity class $\CC$ lying in the polynomial hierarchy. Moreover, we introduce the new notions of strong $\CC$-immunity and strong $\CC$-simplicity. Recall that $\sigmah{k}$-m-quasireductions are all single-valued partial  functions in $\sigmah{k}\mathrm{SV}$ for each $k\in\nat^{+}$.

\begin{definition}\label{def:strong}
Let $\CC\in\{\deltah{k},\sigmah{k}\mid k \in\nat\}$. 
\vs{-2}
\begin{enumerate}
\item A set $S$ is {\em strongly $\CC$-immune} if 
(i) $S$ is infinite and
(ii) for every set $B$ and for every 
$\CC$-m-quasireduction $f$ from $S$
to $B$, $f$ is almost one-to-one on $S$.
\vs{-2}
\item A set $S$ is {\em strongly $\CC$-bi-immune} if $S$ and
$\overline{S}$ are both strongly $\CC$-immune.
\vs{-2}
\item A set $S$ is {\em strongly $\CC$-simple} 
if $S$ is in $\CC$ and 
$\overline{S}$ is strongly $\CC$-immune.
\end{enumerate}
\end{definition} 

In other words, a set $S$ is strongly $\CC$-immune if and only if 
$S$ is infinite and the set $\{ x \in \dom
(f) \mid f(x)=f(u) \} $ is a singleton for every set $B$, for every
$\CC$-m-quasireduction $f$ from $S$ to $B$, and for all but
finitely many strings $u \in \dom (f) \cap S$. In particular, when $\CC=\p$, Definition \ref{def:strong}(2) coincides with the notion of
$\p$-bi-immune sets given by Balc{\'a}zar and Sch{\"o}ning \cite{BS85}.

It directly follows from Lemma \ref{lemma:finite-to-one}(1) that strong
$\CC$-immunity and strong $\CC$-simplicity are indeed restrictions of
$\CC$-immunity and $\CC$-simplicity, respectively. 

\begin{lemma}\label{lemma:strong-to-immune}
For any complexity class $\CC\in\{\deltah{k},\sigmah{k}\mid
k\in\nat\}$, every strongly $\CC$-immune set is $\CC$-immune and every
strongly $\CC$-simple set is $\CC$-simple.
\end{lemma}

A major difference between $\CC$-immunity and strong $\CC$-immunity is shown in the following example. Assume that $\CC$ is closed under $\p$-m-reductions. For any $\CC$-immune set $A$, the disjoint union $A\oplus A$ is also $\CC$-immune; on the contrary, $A\oplus A$ is not strongly $\CC$-immune because it can be reduced to $A$ by the almost two-to-one function $f$ defined by
$f(\lambda)=\lambda$ and $f(xb)=x$ for any $b\in\{0,1\}$, where $\lambda$ is the empty string. Therefore, the class of all strongly $\CC$-immune sets is not closed under the disjoint-union operator. Historically, using the structural difference between these two notions, Balc{\'a}zar and Sch{\"o}ning \cite{BS85} constructed a set in $\e$ which is $\p$-bi-immune but not strongly $\p$-bi-immune. 

What features characterize a set being strongly $\CC$-immune? 
Using a slightly different notion, called a strongly self-bi-immune sets, Balc{\'a}zar and Mayordomo \cite{BM95}
characterized the resource-bounded generic sets of Ambos-Spies, Fleischack, and Huwig \cite{AFH87,AFH88}. Along a similar line of
the study of genericity, we prove the following proposition. 

\begin{proposition}\label{prop:generic-strong}
Let $\CC\in\{\deltah{k},\sigmah{k}\mid k\in\nat\}$. 
Any $\CC$-generic set is strongly $\CC$-bi-immune.
\end{proposition}

\begin{proof}
We show only the case where $\CC=\deltah{k}$ for any fixed number
$k\in\nat^{+}$. Assume that $A$ is $\deltah{k}$-generic but not
strongly $\deltah{k}$-immune. Since $A$ is not strongly
$\deltah{k}$-immune, there exist a set $B$ and a
$\deltah{k}$-m-quasireduction $f$ from $A$ to $B$ such that the collision set $D=\{(x,y)\in\dom(f)^2 \mid x<y \:\wedge\: f(x)=f(y)\}$ is infinite. We denote by $S$ the collection of all forcing conditions $\sigma$ such that there exist at least two elements
$x,y\in\dom(\sigma)\cap\dom(f)$ satisfying that $\sigma(x)\neq
\sigma(y)$ and $f(x)=f(y)$. Since $f$ is in $\fdeltah{k}$, $S$
belongs to $\deltah{k}$.

Next, we show that $S$ is dense. Let $\sigma$ be any forcing
condition. Take a pair $(x,y)$ of distinct strings from the difference 
$\dom(f)\setminus \dom(\sigma)$ that satisfy $f(x)=f(y)$. Such a pair clearly exists because $\dom(\sigma)$ is finite and $D$ is infinite. We define $\tau$ as the unique forcing condition that satisfies the following: $\dom(\tau)=\dom(\sigma)\cup\{x,y\}$, $\sigma\subseteq\tau$, and $\tau(x)\neq\tau(y)$. Obviously, $\tau$ belongs to $S$. Hence, $S$ is indeed dense. Since $A$ is $\deltah{k}$-generic, $A$ must meet $S$; namely, there exists a forcing condition $\rho$ in $S$ such that $A$ extends $\rho$. It thus follows that $f(x')=f(y')$ and $A(x')\neq A(y')$ for a certain pair $(x',y')\in\dom(f)^2$. This contradicts our assumption that $f$ $\deltah{k}$-m-reduces $A$ to $B$. Therefore, $A$ is strongly
$\deltah{k}$-immune. In a similar way, we can also prove that $\overline{A}$ is strongly $\deltah{k}$-immune.
\end{proof}

Now, we consider a closure property of the class of all strongly $\sigmah{k}$-immune sets. We prove that this class is closed under h-$\deltah{k}$-1-reductions. Nevertheless, we cannot replace
h-$\deltah{k}$-1-reductions by h-$\deltah{k}$-m-reductions because the quasireductions that define strong immunity are almost one-to-one. 

\begin{proposition}\label{prop:strong-immune-reduction}
Let $k$ be any number in $\nat^{+}$.
\vs{-2}
\begin{enumerate}
\item The class of all strongly $\sigmah{k}$-immune sets is closed downward under h-$\deltah{k}$-1-reductions on infinite sets.
\vs{-2}
\item The class of all strongly $\sigmah{k}$-immune sets is not closed downward under h-$\p$-m-reductions on infinite sets.
\end{enumerate}
\end{proposition}

\begin{proof}
1) Let $A$ be any infinite set and assume that $A$ is not strongly
$\sigmah{k}$-immune. Assume also that $f$ is an h-$\deltah{k}$-1-reduction from $A$ to a set $B$ and $p$ is a polynomial that witnesses the honesty of $f$. Our goal is to prove that $B$ is not strongly $\sigmah{k}$-immune.

Since $A$ is not strongly $\sigmah{k}$-immune, there exists a set $C$
and a $\sigmah{k}$-m-quasireduction $g$ from $A$ to $C$ such that, for infinitely many strings $u$ in $\dom (g) \cap A $, the set $ \{ x \in \dom (g)\mid g(x)=g(u) \}$ has at least two elements. For readability, write $D$ for the domain of $g$. Hence, $D$ is an infinite $\sigmah{k}$-set. By the definition of $g$, it follows that, for any string $x \in D$, $x \in A$ iff $g(x) \in C$.  Since $f$ is honest, $f(D)$ is an infinite set. Moreover, $f(D)$ belongs to $\sigmah{k}$ because a string $y$ belongs to $f(D)$ if and only if there exists an $x$ such that $|x| \leq p(|y|)$, $x \in D$, and $f(x)=y$.

Next, we define the single-valued partial function $h$ from $B$ to $C$ as follows. Let the domain of $h$ be exactly $f(D)$. For each $y \in \dom(h)$, define $h(y)$ to be $g(x)$, where $x$ is the string in $D$ satisfying that $|x| \leq p(|y|)$ and $f(x)=y$. Since $f$ is one-to-one, such $x$ uniquely exists. Clearly, $h$ is a $\sigmah{k}$-m-quasireduction from $B$ to $C$. Notice that, for infinitely many strings $v$ in $\dom (h) \cap B $, the set $\{ y \in \dom (h) \mid h(y)=h(v) \}$ also has at least two elements. In other words, $B$ is not strongly $\sigmah{k}$-immune, as required.

2) Take any strongly $\sigmah{k}$-immune set $A$. As we discussed just after Lemma \ref{lemma:strong-to-immune}, $A\oplus A$ cannot be strongly $\sigmah{k}$-immune. However, the function $f$ defined by
$f(\lambda)=\lambda$ and $f(xb)=x$ for any $b\in\{0,1\}$ is clearly an h-$\p$-m-reduction from $A\oplus A$ to $A$.  
\end{proof}

As an immediate consequence of Proposition
\ref{prop:strong-immune-reduction}(1), we can show that no strongly
$\sigmah{k}$-simple set can be h-$\deltah{k}$-1-complete for
$\sigmah{k}$, where $k\in\nat^{+}$. The proof of this claim is similar to that of Corollary \ref{cor:reduction-immune} and is left to the avid reader.

\begin{corollary}
For each $k\in\nat^{+}$, no strongly $\sigmah{k}$-simple set is
h-$\deltah{k}$-1-complete for $\sigmah{k}$.
\end{corollary}

We turn our interest to the relativization of strongly $\sigmah{k}$-immune sets. Before giving our main result, we describe a useful lemma that connects $\sigmah{k}$-immunity to strongly $\sigmah{k}$-immunity using the new tower of 2s defined as follows. Let $\hat{2}_0=1$ and let $\hat{2}_n$ be the tower of $2n$ $2$'s, that is, $\hat{2}_{n+1}= 2^{2^{\hat{2}_n}}$ for each $n\in\nat^{+}$. In other words, $\hat{2}_n=\log\log{\hat{2}_{n+1}}$. Let $\hat{T}=\{\hat{2}_n\mid n\in\nat\}$.

\begin{lemma}\label{gap-strong}
Let $\CC\in\{\deltah{k},\sigmah{k}\mid k\in\nat^+\}$ and let $A$ be any set in $\dexp$. If $A$ is $\CC$-immune and $A\subseteq \{1^n\mid n\in\hat{T}\}$, then $A$ is also strongly $\CC$-immune.
\end{lemma}

Note that Lemma \ref{gap-strong} relativizes. 

\begin{proofof}{Lemma \ref{gap-strong}}
We show the lemma only for the case where $\CC=\sigmah{k}$ for any fixed number $k\in\nat^{+}$. Let $A$ be any subset of $\{1^n\mid n\in \hat{T}\}$ in $\dexp$. Assume that $A$ is not strongly $\sigmah{k}$-immune; namely, there exist a $\sigmah{k}$-m-quasireduction $f$ from $A$ to a certain set $B$ such that $f$ is not almost one-to-one on $A$. Since $A\in\dexp$, we can take a deterministic TM $M$ that recognizes $A$ in time at most $2^{n^c+c}$ for a certain constant $c>0$, where $n$ is the length of an input. Let $d$ be the minimal positive integer satisfying $(\log\log{n})^c\leq d\cdot \log{n} +d$ for any number $n\geq0$. We want to show that $A$ is not $\sigmah{k}$-immune.

Fix an element $b$ in $B$. We define the partial function $g$ as follows. On input $x$, if $x\not\in\{1^n\mid \hat{T}\}$, then let $g(x)=f(x)$. Now, assume that $x=1^{\hat{2}_i}$ for a certain number $i\in\nat$. If there exists a natural number $j<i$ such that $f(x)=f(1^{\hat{2}_{j}})$ and $M(1^{\hat{2}_j})=1$, then let $g(x)=b$; otherwise, let $g(x)=f(x)$.  Since $|1^{\hat{2}_j}|=\hat{2}_j\leq \log\log{|x|}$, the running time of $M$ on input $1^{\hat{2}_j}$ is at most $2^{(\log\log{|x|})^c+c}$, which is further bounded above by $2^{c+d}|x|^d$ for any nonempty string $x$. Hence, $g$ is in $\sigmah{k}\mathrm{SV}$. Consider the set $g^{-1}(b)$. Since $f$ is not almost one-to-one, $g$ maps infinitely many strings in $A$ into $b$. Thus, $g^{-1}(b)$ must be infinite. Obviously, $g$ $\sigmah{k}$-m-quasireduces $A$ to $B$. By Lemma \ref{lemma:finite-to-one}(1), $A$ is not $\sigmah{k}$-immune.
\end{proofof}

We show in Proposition \ref{strong-oracle} a relativized world where strongly $\sigmah{k}$-simple sets actually exist. Our proof of this proposition heavily relies on Bruschi's \cite{Bru92} construction of a recursive oracle relative to which a $\sigmah{k}$-simple set exists. 

\begin{proposition}\label{strong-oracle}
For each $k\in\nat^{+}$, there exists a strongly $\sigmah{k}(A)$-simple set relative to a certain recursive oracle $A$.
\end{proposition}

\begin{proof}
Let $k\in\nat^{+}$. For each oracle $A$, we define the oracle-dependent set:
\[
L_k^A = \{1^n \mid n\in \hat{T}\:\wedge\: \forall y_1\in\Sigma^{n}\exists y_2\in\Sigma^{n}\cdots Q_k y_k\in\Sigma^{n}[1^ny_1y_2\cdots y_k\not\in A]\},
\]
where $Q_k$ is $\exists$ if $k$ is even and $Q_k$ is $\forall$ otherwise. Obviously, $L_k^A$ is in $\pih{k}(A)$ for any oracle $A$. The key to our proof is the existence of a recursive oracle $A$ that makes $L^A_k$ $\sigmah{k}(A)$-immune. In his proof of \cite[Theorem 5.4]{Bru92}, Bruschi employed a well-studied circuit lower bound technique in the course of the construction of an immune set. We do not attempt to recreate his proof here; however, we note that his construction works on any sufficiently large string input, in particular, of the form $1^m$. Therefore, we can build in a recursive fashion an oracle $A$ for which $L_k^A$ is $\sigmah{k}(A)$-immune. Since $L_k^A$ is obviously in $\dexp^A$, Lemma \ref{gap-strong} ensures that $L_k^A$ is strongly $\sigmah{k}(A)$-immune. 
\end{proof}

Concerning a random oracle, Vereshchagin \cite{Ver93,Ver95} earlier demonstrated the existence of an $\np$-simple set relative to a random oracle. By analyzing his construction in \cite{Ver95}, we can prove the existence of strongly $\np$-simple sets relative to a random oracle.

\begin{proposition}\label{strong-random}
Relative to a random oracle, a strongly $\np$-simple set exists.
\end{proposition}

\begin{proof}
In \cite[Theorem 3]{Ver93}, Vereshchagin proved that the oracle-dependent tally set: 
\[
L^X=\{1^n\mid n\in
Tower \:\wedge\: \forall w\in\Sigma^n \exists v\in\Sigma^{\floors{\log n}}[wv\in A]\}
\]
has no $\np^X$-subsets relative to a random oracle $X$. For our purpose, we further define the set $K^A= \{1^n\mid n\in \hat{T} \:\wedge\: 1^n\in L^A\}$. Clearly, $K^A$ is in $\co\np^A$ for any oracle $A$. Since $K^A\subseteq\{1^n\mid n\in \hat{T}\}$ and $K^A\in\dexp^A$, by Lemma \ref{gap-strong}, if $K^A$ is $\np^A$-immune then it is also strongly $\np^A$-immune. Similar to Vereshchagin's proof, we can show that $K^X$ is $\np^X$-immune relative to a random oracle $X$. Therefore, $K^X$ is strongly $\np^X$-immune relative to a random oracle $X$, as requested. 
\end{proof}

Finally, we prove in Theorem \ref{theorem:strongly-simple} that a strongly $\np^G$-simple set exists relative to a Cohen-Feferman generic oracle $G$. This immediately implies the existence of an $\np^G$-simple set relative to the same generic oracle $G$. 

\begin{theorem}\label{theorem:strongly-simple}
A strongly $\np^G$-simple set exists relative to a Cohen-Feferman
generic oracle $G$.
\end{theorem}

The proof of Theorem \ref{theorem:strongly-simple} uses {\em weak forcing} instead of Feferman's original {\em finite forcing}. Let $\varphi(X)$ be any arithmetical statement including variable $X$, which runs over all subsets of $\Sigma^*$. We say that a forcing condition $\sigma$ {\em forces} $\varphi$ (notationally, $\sigma\forces\varphi(X)$) if $\varphi(G)$ is true for every Cohen-Feferman generic set $G$ that extends $\sigma$. This forcing relation $\forces$ satisfies the following five properties: 
\begin{enumerate}
\item $\sigma\forces \neg\:\varphi$ $\IFF$ 
no extension of $\sigma $ forces $\varphi$, 
\vs{-2}
\item $\sigma\forces (\varphi \wedge \psi)$ $\IFF$ 
$\sigma\forces \varphi $ and $\sigma\forces \psi$, 
\vs{-2}
\item $\sigma\forces (\varphi \vee \psi)$ $\Longrightarrow$ 
$\sigma $ has an extension $\rho$ such that 
$\rho\forces \varphi $ or $\rho\forces \psi$, 
\vs{-2}
\item $\sigma\forces \forall\: x\: \varphi (x)$ $\IFF$ 
$\sigma\forces \varphi (n)$ for all instances $n$, and
\vs{-2}
\item $\sigma\forces \exists\: x\: \varphi (x)$ $\Longrightarrow$ 
$\sigma $ has an extension $\rho$ such that 
 $\rho\forces \varphi (n)$ for a certain instance $n$. 
\end{enumerate}
In properties 4 and 5, the symbol $n$ represents either a natural number or a string. Notice that properties 3 and 5 have only one-way implications. See a standard textbook of, \eg Odifreddi \cite{Odi89} for more details on weak forcing and its connection to Feferman's finite forcing. Note that, in Odifreddi's book, Cohen-Feferman generic sets are called {\lq\lq}$\omega$-generic{\rq\rq} sets. In our setting, the domain of a forcing condition may be any finite set of strings; thus, a forcing condition is not necessarily an initial segment of an oracle.  In the proof of Theorem \ref{theorem:strongly-simple}, a {\lq\lq}forcing relation{\rq\rq} always refers to {\em weak forcing}.

For simplicity, we encode a computation path of a nondeterministic oracle TM into a binary string and identify such a path with its encoding. Using this encoding, we can enumerate lexicographically all the computation paths of the machine. For any such machine $N$, any oracle $B$, and any string $x$, we define the useful set $Q_{\boxplus}(N,B,x)$ as follows. When $N^B$ rejects input $x$, $Q_{\boxplus} (N,B,x)$ is set to be empty. Assume otherwise. Take the lexicographically first accepting computation path $\gamma$ of $N^B$ on input $x$ and let $Q_{\boxplus} (N,B,x)$ be the set of all strings queried along this computation path $\gamma$. 

\begin{proofof}{Theorem \ref{theorem:strongly-simple}}
In this proof, we use the oracle-dependent set $L^X = \{ x \mid \forall y\in\Sigma^{|x|} [ xy \not\in X ] \}$, where $X$ is any oracle. Letting $G$ be any Cohen-Feferman generic set, 
we wish to show that $L^G$ is strongly $\np^G$-immune. This implies that $\overline{L^G}$ is strongly $\np^G$-simple since $\overline{L^G}$ is in $\np^G$. 

Let $k$ be any number in $\nat^{+}$. Let $N$ be any nondeterministic oracle TM with an output tape running within time $n^k+k$ independent of the choice of an oracle. Henceforth, we use the symbol $X$ to denote a variable running over all subsets of $\Sigma^*$. We define $f^X$ to be the function computed by $N^X$. In general, $f^X$ is a multi-valued function. By the choice of $N$, on any input $x$, $N^X$ always enters an accepting state or a rejecting state. Thus, by our convention, whenever $N^X$ produces no accepting computation path on input $x$, $f^X(x)$ is undefined. Now, assume that $f^G$ is an $\np^G$-m-quasireduction mapping from $L^G$ to a certain set. Consider the  set: 
\[
K^X = \{ (x, y) \in \dom (f^X)^2 \mid x\in L^X \:\wedge\: 
x < y \:\wedge\: f^X(x)=f^X(y) \}. 
\]
Since $f^G$ is a $\np^G$-m-quasireduction from $L^G$, the set $K^G$ coincides with the original collision set defined in the beginning of this section. It therefore suffices to show that $K^G$ is finite. Because of the definition of weak forcing, we can assume without loss of generality that the following three statements are forced by the empty forcing condition (in other words, they are forced by every forcing condition). 
\begin{itemize}
\item[1)] For any string $x$, 
$x \not\in \dom (f^X) \IFF N^X$ on input $x$ enters no accepting state, 
\vs{-2}
\item[2)] For any string $x\in\dom(f^X)$, 
$N^X$ outputs $f^X(x)$ on {\em all}\/ accepting computation paths, 
\vs{-2}
\item[3)] For any $n\in \nat$, $N^X$ runs within time $n^k + k$ on every input of length $n$.
\end{itemize}
Now, consider the following arithmetical statements, which intuitively state that every element in $K^X$ is bounded above by a certain constant $n$. 

\begin{itemize}
\item $\varphi_0(X) {\equiv}$ $\exists\,(u,v)\in(\Sigma^*)^2\: [f^X(u) = f^X(v) \:\wedge\: u\in L^X\:\wedge\: v\not\in L^X]$, and 
\vs{-2}
\item ${\varphi}_1(X) {\equiv}$ 
$\exists\, n \in \omega\: \forall\, (u,v) \in (\Sigma^*)^2\:
[u,v\not\in\dom(f^X) \:\vee\: u\not\in L^X \:\vee\: u\geq v \:\vee\: f^X(u)\neq f^X(v) \:\vee\: |v| \leq n]$. 
\end{itemize}

\n For simplicity, write $\varphi(X)$ for $(\varphi_0(X)\:\vee\:\varphi_1(X))$. Notice that, in general, $f^A$ might not $\np^A$-m-quasireduce $L^A$ for a certain oracle $A$.  Our goal is to prove that $\varphi(G)$ is true. This clearly implies that $K^G$ is finite since $\varphi_0(G)$ is obviously false and thus $\varphi_1(G)$ must be true. To achieve our goal, we define $\DD=\{\sigma\mid \sigma\forces \varphi(X)\}$ and then claim that $\DD$ is dense. Assuming that $\DD$ is dense, the genericity of $G$ guarantees the existence of a forcing condition $\sigma$ such that $\sigma\forces \varphi(X)$ and $\sigma\subseteq G$. By the definition of weak forcing, $\varphi(G)$ must be true. This will complete the proof.

It is thus enough to prove that $\DD$ is dense. Let $\sigma $ be any forcing condition. We want to show that there exists an extension $\tau$ of $\sigma$ in $\DD$. Assume otherwise that no extension of $\sigma$ forces $\varphi(X)$. {}From property 1 of weak forcing, $\sigma$ forces $\neg\:\varphi(X)$, which implies $\sigma\forces\neg\:\varphi_0(X)$ and $\sigma\forces\neg\:\varphi_1(X)$. Since $\sigma\forces\neg\:\varphi_1(X)$, $\sigma$ forces the statement:
\[
\forall\, n \in \omega \: \exists\, (u,v) \in (\Sigma^*)^2 [u,v\in\dom(f^X)\:\wedge\: u\in L^X \:\wedge\: u<v \:\wedge\: f^X(u)=f^X(v) \:\wedge\: |v| > n].
\]
Take any natural number $n$ that is greater than $|\dom(\sigma)|$. By properties 4 and 5, there exist a forcing condition $\rho$ extending $\sigma$ and a pair $(u,v)$ of strings  such that $|v| > n$, $u < v$, and $\rho$ forces {\lq\lq}$u,v \in \dom (f^X) \:\wedge\: u\in L^X \:\wedge\: f^X(u)=f^X(v)$.{\rq\rq} In what follows, since the aforementioned statements 1), 2), and 3) are forced by $\rho$, we can assume that the domain of $\rho$ consists only of the following four sets: (i) $\dom(\sigma)$, (ii) $u\Sigma^{|u|}$  (to force {\lq\lq}$u \in L^X${\rq\rq}), (iii) $Q_{\boxplus}(N,\rho,u)$ (to decide the value of $f^{X}(u)$), and (iv) $Q_{\boxplus}(N,\rho,v)$ (to decide the value of $f^X(v)$). Note that $|Q_{\boxplus}(N,\rho,u)|\leq |u|^k + k$ and $|Q_{\boxplus}(N,\rho,v)|\leq |v|^k + k$. Note also that $u\Sigma^{|u|}$ and $v\Sigma^{|v|}$ are disjoint and $|v| \geq \max \{ |\dom (\sigma)|, |u| \} $. Since $u<v$ and $|v| > n$, the cardinality $|v\Sigma^{|v|} \cap \dom (\rho)|$ is at most $2(|v|^k + k) + |v|$.  Therefore, there exists at least one string $w$ 
of length $|v|$ such that $vw \not\in \dom (\rho)$. 
With this $w$, we can extend $\rho$ to another forcing condition $\tau$ that forces {\lq\lq}$u \in L^X \:\wedge\: v \not\in L^X \:\wedge\: f^X(u)=f^X(v)$.{\rq\rq} This contradicts the assumption that $\sigma\forces\neg\:\varphi_0(X)$. Consequently, ${\cal D}$ is dense. This completes the proof of the theorem. 
\end{proofof}

\section{Almost Immunity and Almost Simplicity}
\label{sec:almost-immune}

We have shown in the previous section that strong $\CC$-immunity and strong $\CC$-simplicity strengthen the ordinary notions of $\CC$-immunity and $\CC$-simplicity. In contrast to these notions, Orponen \cite{Orp86} and Orponen, Russo, and Sch{\"o}ning  \cite{ORS86} extended $\p$-immunity to the notion of almost $\p$-immunity. The complementary notion of almost $\p$-immunity under the term $\p$-levelability (a more general term {\lq\lq}levelable sets{\rq\rq} was first used by Ko \cite{Ko85} in a resource-bounded setting) was extensively discussed by Orponen \etalc~\cite{ORS86}.  Naturally, we can generalize these notions to almost $\CC$-immunity and $\CC$-levelability for any complexity class $\CC$. Furthermore, we  introduce the new notions of almost $\CC$-bi-immune sets and almost $\CC$-simple sets.

\begin{definition}\label{def:almost-immune}
Let $\CC$ be any complexity class.
\vs{-2}
\begin{enumerate}
\item A set $S$ is {\em almost $\CC$-immune} if $S$ is the union of  
a $\CC$-immune set and a set in $\CC$.
\vs{-2}
\item An infinite set is {\em $\CC$-levelable} iff it is not almost
$\CC$-immune.
\vs{-2}
\item A set $S$ is {\em almost $\CC$-bi-immune} if $S$ and $\overline{S}$
are both almost $\CC$-immune.
\vs{-2}
\item A set $S$ is {\em almost $\CC$-simple} if $S$ is an infinite set in
$\CC$ and $\overline{S}$ is the union of a set $A$ in $\CC$ and a
$\CC$-immune set $B$, where $B\setminus A$ is infinite.
\end{enumerate}
\end{definition}

It follows from Definition \ref{def:almost-immune}(1) that every almost $\CC$-immune set is infinite. The definition of almost $\CC$-simplicity in Definition
\ref{def:almost-immune}(4) is slightly different from other simplicity definitions because the infinity condition of the difference $B\setminus A$ is necessary to guarantee $\CC\neq\co\CC$, provided that an almost $\CC$-simple set exists. This is shown in the following lemma.

\begin{lemma}
Let $\CC$ be any complexity class closed under finite variations,
finite union and finite intersection. If an almost $\CC$-simple set
exists, then $\CC\neq \co\CC$.
\end{lemma}

\begin{proof}
Assume that $\CC=\co\CC$ and let $S$ be any almost $\CC$-simple
set. There exist a set $A \in \CC$ and a $\CC$-immune set $B$ such
that $\overline{S}=A\cup B$ and $B\setminus A$ is infinite. Since
$\overline{S},\overline{A}\in\CC$, $\overline{S}\setminus A$ is in $\CC$. Note
that $B\setminus A\subseteq B$ and $B\setminus A=\overline{S}\setminus A \in \CC$. Since $B$ is $\CC$-immune, $B\setminus A$ must be finite. This contradicts the almost $\CC$-immunity of $S$. Therefore, the lemma holds.
\end{proof}

The following lemma is immediate from Definition \ref{def:almost-immune}.

\begin{lemma}\label{lemma:immune-almost}
For any complexity class $\CC$, every $\CC$-immune set is almost
$\CC$-immune and every $\CC$-simple set is almost $\CC$-simple.
\end{lemma}

Several characterizations of almost $\p$-immunity and $\p$-levelability are shown by Orponen \etalc~\cite{ORS86} in terms of maximal $\p$-subsets and polynomial-time $\p$-to-finite reductions. We can naturally extend these characterizations to almost $\deltah{k}$-immunity and $\deltah{k}$-levelability (but not to the $\Sigma$-level classes of the polynomial hierarchy). We leave such an extension to the avid reader.

To understand the characteristics of almost $\CC$-immunity, we begin
with a simple observation. We say that a set $S$ is {\em polynomially paddable} ({\em paddable}, in short) if there is a one-to-one polynomial-time computable function $pad$ (called the {\em padding function}) from $\Sigma^{*}$ to $\Sigma^{*}$ such that, for all pairs $x,y\in\Sigma^{*}$, $x\in S$ iff $pad(\pair{x,y})\in
S$. A set $S$ is {\em honestly paddable} if it is paddable with a
padding function that is componentwise honest. It is proven by Orponen \etalc~\cite{ORS86} that any honestly paddable set not in $\p$ is $\p$-levelable. As observed by Russo \cite{Rus86}, the essence of this assertion is that if $A\not\in\p$ and $A$ is length-increasing $\p$-m-autoreducible then $A$ is $\p$-levelable, where $A$ is {\em length-increasing $\CC$-m-autoreducible} if $A$ is $\CC$-m-reducible to $A$ via a certain length-increasing reduction. This observation can be generalized to $\deltah{k}$-levelable sets in the following lemma. The second part of the lemma will be used in Section \ref{sec:hyperimmune}. 

\begin{lemma}\label{lemma:paddable}
Let $k\in\nat^{+}$ and $A\subseteq\Sigma^*$. Assuming that $A\not\in\deltah{k}$, if $A$ is length-increasing $\deltah{k}$-$m$-autoreducible, then $A$ and $\overline{A}$ are both $\deltah{k}$-levelable. Thus, if $\deltah{k}\neq\sigmah{k}$ then $\sigmah{k}$ as well as $\pih{k}$ has a $\deltah{k}$-levelable set.
\end{lemma}

Although our proof is a simple extension of Russo's \cite{Rus86}, we include it for completeness. For any function $f$ and any number $k\in\nat^{+}$, the notation $f^{(k)}$ denotes the $k$-fold composition of $f$. Note that $f^{(1)}=f$. For convenience, we define $f^{(0)}$ to be the identity function.

\begin{proofof}{Lemma \ref{lemma:paddable}}
Assume to the contrary that $A$ is almost $\deltah{k}$-immune and $A$ is $\deltah{k}$-m-autoreducible via a length-increasing reduction $f$. Since $\overline{A}$ is also $\deltah{k}$-m-autoreducible via $f$, it suffices to prove the lemma only for $A$. Assume also that $A$ is outside of $\deltah{k}$. Since $A$ is almost $\deltah{k}$-immune, $A$ is expressed as of the form $B\cup C$, where $B$ is a set in $\deltah{k}$ and $C$ is a $\deltah{k}$-immune set. Note that the difference $C\setminus B$ is infinite since, otherwise, $A$ falls into $\deltah{k}$. Now, set $D=\{x\mid x\not\in B\:\wedge\: f(x)\in B\}$. Clearly, $D\subseteq C$.

To lead to a contradiction, it suffices to show that $D$ is an infinite set in $\deltah{k}$. Since $f$ is in $\fdeltah{k}$, $D$ belongs to $\deltah{k}$. If $D$ is finite, then the difference $C\setminus (B\cup D)$ must be infinite. In this case, take the lexicographically largest element $z_0$ in $D$ and also take the minimal string $x$ in $C\setminus (B\cup D)$ such that $|x|>|z_0|$. The set $F=\{f^{(i)}(x)\mid i\in\nat\}$ must include an element in $B$ because, otherwise, $F$ becomes an infinite $\deltah{k}$-subset of $C$, a contradiction. Hence, there exists a number $k\in\nat$ such that $f^{(k)}(x)$ falls in $D$. This implies that $|f^{(k)}(x)|\leq|z_0|<|x|$, which contradicts the length-increasing property of $f$. Therefore, $D$ is infinite, as required.

The second part of the lemma follows from the fact that, under the assumption $\deltah{k}\neq\sigmah{k}$, the class $\sigmah{k}$ as well as $\pih{k}$ contains honestly paddable sets not in $\deltah{k}$, which satisfy the premise of the first part of the lemma.
\end{proofof}

Most known $\p$-m-complete sets for $\np$ are known to be honestly paddable
and thus, by Lemma \ref{lemma:paddable}, the complements of these sets are $\p$-levelable sets, which are also $\np$-levelable, unless $\p=\np$. Therefore, most known $\p$-m-complete sets for $\np$ cannot be almost $\np$-simple. This result can be compared with Proposition \ref{almost-complete}.

Earlier, Ko and Moore \cite{KM81} considered the resource-bounded notion of {\lq\lq}productive sets.{\rq\rq}  Another formulation based on $\np_{(k)}$ was later given by Joseph and Young \cite{JY85}, who used the terminology of {\lq\lq}$k$-creative sets,{\rq\rq} where $k$ is any number in $\nat^{+}$. Now, fix $k\in\nat^{+}$. A set $S$ in $\np$ is called {\em $k$-creative} if there exists a function $f\in\fp$ such that, for all $x\in \gindex_{(k)}$, $f(x)\in S$ iff $f(x)\in W_{x}$. This function $f$ is called the {\em productive function} for $S$. If in addition $f$ is honest, $S$ is called {\em honestly $k$-creative}.  Joseph and Young showed that every $k$-creative set is $\p$-m-complete for $\np$. Later, Orponen \etalc~\cite{ORS86} showed that, unless $\p=\np$, every honestly $k$-creative set is $\p$-levelable by demonstrating that any honestly $k$-creative set is length-increasing $\p$-m-autoreducible.  {}From Lemma \ref{lemma:paddable}, it follows that if $\p\neq \np$ then any honestly $k$-creative set and its complement are both $\p$-levelable. Consequently, we obtain the following result.

\begin{corollary}
For any $k\in\nat^{+}$, no honestly $k$-creative set is almost $\np$-simple.
\end{corollary}

Our notion of almost $\CC$-simplicity is similar to what Uspenskii
\cite{Usp57} discussed under the term
{\lq\lq}pseudosimplicity.{\rq\rq} Here, we give a resource-bounded
version of his pseudosimplicity. A set $S$ is called {\em
$\CC$-pseudosimple} if there is an infinite $\CC$-subset $A$ of
$\overline{S}$ such that $S\cup A$ is $\CC$-simple. Although $\CC$-simple sets cannot be $\CC$-pseudosimple by our
definition,  any infinite $\CC$-pseudosimple set is almost $\CC$-simple. The latter claim is shown as follows.  Suppose that $S$ is an infinite $\CC$-pseudosimple set and $A$ is a $\CC$-subset of $\overline{S}$ for which $S\cup A$ is $\CC$-simple. This means that $\overline{S}\setminus A$ is $\CC$-immune. Therefore, $S$ is almost $\CC$-simple.

The following theorem shows a close connection among simplicity,
almost simplicity, and pseudosimplicity. This theorem signifies the
importance of almost simple sets.  

\begin{theorem}\label{theorem:pseudosimple}
For each $k\in\nat^{+}$, 
the following three statements are equivalent. 
\vs{-2}
\begin{enumerate}
\item There exists a $\sigmah{k}$-simple set. 
\vs{-2}
\item There exists an infinite $\sigmah{k}$-pseudosimple set in
$\p$.
\vs{-2}
\item There exists an almost $\sigmah{k}$-simple set in 
$\p$. 
\end{enumerate}
\end{theorem}

\begin{proof}
Let $k$ be any number in $\nat^{+}$. 

2 implies 3) This implication holds because any infinite $\sigmah{k}$-pseudosimple set is indeed almost $\sigmah{k}$-simple as mentioned before.

3 implies 1) Assume that $S$ is an almost $\sigmah{k}$-simple set in
$\p$. By definition, there exist a set
$A$ in $\sigmah{k}$ and a $\sigmah{k}$-immune set $B$ such that
$\overline{S}=A\cup B$ and $B\setminus A$ is infinite. Define $C =B\setminus A$. We show that $\overline{C}$ is the desired $\sigmah{k}$-simple set. First, since $\overline{C}=S\cup A$ and $A\in\sigmah{k}$, $\overline{C}$ belongs to $\sigmah{k}$. Second, since $C\subseteq B$, $C$ is $\sigmah{k}$-immune. This yields the $\sigmah{k}$-simplicity of $\overline{C}$.

1 implies 2) Suppose that there exists a $\sigmah{k}$-simple set
$S$. Under this assumption, we want to claim that both $0\Sigma^*$ and
$1\Sigma^*$ are $\sigmah{k}$-pseudosimple.  This is shown as
follows. For each bit $a\in\{0,1\}$, let $A_{a} = a S$
and $B_{a} = a\overline{S}$. The immunity of $\overline{S}$ implies that $B_{0}$ and $B_{1}$ are both infinite. Consider the case for $0\Sigma^*$. Note that $A_1$ is a $\sigmah{k}$-subset of $1\Sigma^*$. Since $\overline{S}$ is $\sigmah{k}$-immune and $B_{1}\subseteq 1\overline{S}$, it follows that $B_{1}$ is $\sigmah{k}$-immune. Observe that $B_1=1\Sigma^*\cap \overline{A_1}$. Hence, its complement $0\Sigma^*
\cup A_1$ is a $\sigmah{k}$-simple set. This concludes that $0\Sigma^*$ is $\sigmah{k}$-pseudosimple. By a similar argument, $1\Sigma^*$ is $\sigmah{k}$-pseudosimple.
\end{proof}

Theorem \ref{theorem:pseudosimple} indicates the importance of the structure of $\p$ in the course of the study of $\sigmah{k}$-simplicity. In Bruschi's \cite{Bru92} relativized world where a $\sigmah{k}$-simple set exists, since Theorem \ref{theorem:pseudosimple} relativizes, there also exists an almost $\sigmah{k}$-simple set within $\p$.  

We note the relativization of almost $\np$-simple sets. Vereshchagin  \cite{Ver95} proved that there are two partitions $L_0^X$ and $L_1^X$ of the set $\{1^n\mid n\in Tower\}$ such that $L_0^X$ is $\np^X$-immune and in $\co\np^X$ and $L_1^X$ is $\co\np^X$-immune and in $\np^X$ relative to a random oracle $X$. This implies that the set $L_1^X$ is an almost $\np^X$-simple set that is also $\co\np^X$-immune relative to a random oracle $X$.

Finally, we briefly discuss a closure property of the class of all almost $\sigmah{k}$-immune sets under polynomial-time reductions. For each number $k$ in $\nat^{+}$, the class
of all $\sigmah{k}$-immune sets is closed under
h-$\deltah{k}$-d-reductions on infinite sets whereas the class of all
almost $\sigmah{k}$-immune sets is closed under
h-$\deltah{k}$-m-reductions on infinite sets.
The latter claim is proven as follows. Let $A$ be any infinite set.  Assume that $f$ is an h-$\p$-m-reduction from $A$ to an almost $\sigmah{k}$-immune set $B$. This means that $B$ is the union of two subsets $B_1$ and $B_2$, where $B_1$ is $\sigmah{k}$-immune and $B_2 \in \sigmah{k}$. Therefore, $A$ should be the union of the two subsets $f^{-1} (B_1)$ and $f^{-1}(B_2)$. Since $B_1$ is infinite, $f^{-1} (B_1)$ is also infinite. Hence, by Proposition \ref{theorem:reduction-immune}, $f^{-1} (B_1)$ is $\sigmah{k}$-immune. Since $f$ is honest, it follows that $f^{-1} (B_2) \in \sigmah{k}$. Hence, $A$ is almost $\sigmah{k}$-immune.  This immediately implies the following consequence.

\begin{proposition}\label{almost-complete}
For each $k\in\nat^{+}$, no almost $\sigmah{k}$-simple set is h-$\deltah{k}$-m-complete for $\sigmah{k}$.
\end{proposition}

\section{Hyperimmunity and Hypersimplicity}
\label{sec:hyperimmune}

Since Post \cite{Pos44} constructed a so-called {\em hypersimple} set, the notions of hyperimmunity and hypersimplicity have played a significant role in the progress of classical recursion theory. A resource-bounded version of these notions was first considered by Yamakami \cite{Yam95} and studied extensively by Schaefer and Fenner \cite{SF98}. The definition of Schaefer and Fenner is based on the notion of {\lq\lq}honest $\np$-arrays,{\rq\rq} which differs from the notion of {\lq\lq}strong arrays{\rq\rq} in recursion theory, where a {\em strong array} is a series of pairwise disjoint finite sets. For our formalization, we demand only {\lq\lq}eventual disjointness{\rq\rq} for sets in an array rather than {\lq\lq}pairwise disjointness.{\rq\rq}

A binary string $x$ is said to {\em represent} the finite set $\{a_1,a_2,\ldots,a_k\}$ if and only if 
$x=\pair{a_1,a_2,\ldots,a_k}$ and $a_1<a_2<\cdots<a_k$ in the
standard lexicographic order on $\Sigma^*$. For convenience, we say that a set $A$ {\em surpasses} another set $B$ if there exists a string $z$ in $A$ satisfying that $z>x$ (lexicographically) for all strings $x\in B$.

\begin{definition}
Let $k\in\nat^{+}$, $A\subseteq\Sigma^*$, and  $\CC\in\{\sigmah{k},\deltah{k}\}$. 
\vs{-2}
\begin{enumerate}
\item An infinite sequence $\DD= \{D_{s}\}_{s\in\Sigma^*}$ of finite sets is called a {\em $\sigmah{k}$-array} ({\em $\deltah{k}$-array},
resp.)  if there exists a {\em partial} function $f$ in
$\sigmah{k}\mathrm{SV}$ ($\fdeltah{k}$, resp.) such that (i)
$\dom (f)$ is infinite, (ii) $D_s\neq\setempty$ and $f(s)$ represents $D_{s}$ for any string $s\in\dom(f)$, and (iii) $D_s=\setempty$ for any string $s\not\in\dom(f)$. This $f$ is called a {\em supporting function} of $\DD$ and the set $\bigcup_{s\in\dom(f)}D_s$ is called the {\em support} of $\DD$. The {\em width} of $\DD$ is the supremum of the cardinality $|D_s|$ over all indices $s\in\dom(f)$.
\vs{-2}
\item A $\CC$-array $\DD$ has an {\em
infinite support} if the support of $\DD$ is infinite.
\vs{-2}
\item A $\CC$-array $\{D_{s}\}_{s\in\Sigma^*}$ via $f$ is {\em
polynomially honest} ({\em honest}, in short) if $f$ is componentwise honest; namely, there exists a polynomial $p$ such that $|s|\leq p(|x|)$ for any $s\in\dom(f)$ and any $x\in D_{s}$.
\vs{-2}
\item A $\CC$-array $\{D_{s}\}_{s\in\Sigma^*}$ via $f$ is {\em eventually
disjoint} if, for every string $x\in\dom(f)$, there exists a string
$y$ in $\dom(f)$ such that $y\geq x$ (lexicographically), $D_{x}\cap D_{y}=\setempty$, and $D_y$ surpasses $D_x$.
\vs{-2}
\item A $\CC$-array $\{D_s\}_{s\in\Sigma^*}$ via $f$ {\em intersects} $A$ if $D_s\cap A\neq \setempty$ for any $s$ in $\dom(f)$.
\end{enumerate} 
\end{definition}

The honesty condition of an $\CC$-array guarantees that the array is
eventually disjoint. In addition, any eventually disjoint $\CC$-array has an infinite support because, for any element $D$ in the array, we can always find another element $D'$ which surpasses $D$.

A simple relationship between $\sigmah{k}$-simplicity and a honest $\sigmah{k}$-array is given in the following lemma, which was implicitly proven by Yamakami \cite{Yam95} and later explicitly stated by Schaefer and Fenner \cite{SF98} for the case where $k=1$.

\begin{lemma}
Let $k\in\nat^{+}$. For any $\sigmah{k}$-simple set $A$ and any $\sigmah{k}$-array $\DD$, if $\DD$ intersects $\overline{A}$, then the width of $\DD$ is infinite. 
\end{lemma}

\begin{proof}
Let $A$ be any $\sigmah{k}$-simple set and let $\ell$ be any number in $\nat^{+}$. To lead to a contradiction, we assume that there exists a honest $\sigmah{k}$-array $\DD = \{D_x\}_{x\in\Sigma^*}$ via $f$ such that the width of $\DD$ is $\leq \ell$ and $\DD$ intersects $\overline{A}$. For brevity, write $D$ for the support of $\DD$. Obviously, $D$ is infinite and in $\sigmah{k}$. Since $f$ is componentwise honest, we can take a polynomial $p$ satisfying $|x|\leq p(|y|)$ for all $x\in\dom(f)$ and all $y\in D_x$. Now, define $\ell_{max}$ to be the maximal value $i$ such that $|D_x\cap A|=i$ for infinitely many strings $x$ in $\dom(f)$. Note that the $\sigmah{k}$-simplicity of $A$ guarantees that $\ell_{max}>0$. Take any sufficiently large number $n_0\in\nat$ that guarantees $|D_x\cap A|\leq \ell_{max}$ for all strings $x$ of length $\geq n_0$.

Define the set $B=\{y\mid \exists x[n_0\leq |x|\leq p(|y|)\:\wedge\: y\in D_x \:\wedge\: |(D_x\setminus\{y\})\cap A|=\ell_{max}]\}$, which is clearly in $\sigmah{k}$ because so are $A$ and $D$. This $B$ is infinite because the set $\{x\in\dom(f) \mid |x|\geq n_0\:\wedge\: |D_x\cap A|=\ell_{max}\}$ and $D$ are both infinite and $\DD$ intersects $\overline{A}$. Now, we show that $B\subseteq\overline{A}$. This is true because, if $y\in B\setminus \overline{A}$, then 
$\ell_{max}= |(D_x\setminus\{y\})\cap A| < |D_x\cap A|\leq \ell_{max}$, a contradiction. Therefore, $B$ is an infinite subset of $\overline{A}$. Since $B$ is in $\sigmah{k}$, $\overline{A}$ cannot be $\sigmah{k}$-immune. This contradicts our assumption that $A$ is $\sigmah{k}$-simple.
\end{proof}

We introduce below the notions of $\CC$-hyperimmunity and honest
$\CC$-hyperimmunity. 

\begin{definition}
Let $\CC\in\{\deltah{k},\sigmah{k}\mid k\in\nat\}$.
\vs{-2}
\begin{enumerate}
\item A set $S$ is {\em (honestly) $\CC$-hyperimmune} if $S$ is infinite
and there is no (honest) $\CC$-array $\DD$ such that $\DD$ is
eventually disjoint and intersects $S$.
\vs{-2}
\item A set $S$ is {\em (honestly) $\CC$-bi-hyperimmune} if $S$ and
$\overline{S}$ are both (honestly) $\CC$-hyperimmune.
\vs{-2}
\item A set $S$ is {\em (honestly) $\CC$-hypersimple} if $S$ is in $\CC$
and $\overline{S}$ is (honestly) $\CC$-hyperimmune.
\end{enumerate}
\end{definition}

Note that {\lq\lq}$\np$-hyperimmunity{\rq\rq} defined by Schaefer and Fenner \cite{SF98} coincides with our honest $\np$-hyperimmunity.
Any honestly $\CC$-hyperimmune set is $\CC$-immune because, assuming that $S$ is not $\CC$-immune, we can choose an infinite subset $A$ of $S$ in $\CC$ and define $D_s=\{s\}$ if $s\in A$, and $D_s=\setempty$ otherwise, which shows that $A$ is not honestly $\CC$-hyperimmune.

\begin{lemma}
For any complexity class $\CC\in\{\sigmah{k},\deltah{k}\mid
k\in\nat\}$, every honestly $\CC$-hyperimmune set is $\CC$-immune and
every honestly $\CC$-hypersimple set is $\CC$-simple.
\end{lemma}

In Proposition \ref{prop:generic-strong}, we have seen that $\CC$-generic sets are strongly $\CC$-bi-immune. Similarly, $\CC$-generic sets are examples of honestly $\CC$-hyperimmune sets.

\begin{proposition}\label{generic-hyperimmune}
Let $k\in\nat^{+}$. All $\sigmah{k}$-generic sets are honestly $\sigmah{k}$-hyperimmune.
\end{proposition}

The following proof works for $\sigmah{k}$-genericity but not for $\deltah{k}$-genericity.

\begin{proofof}{Proposition \ref{generic-hyperimmune}}
Fix $k\in\nat^{+}$ and let $A$ be any $\sigmah{k}$-generic set. Assume that $A$ is not honestly $\sigmah{k}$-hyperimmune; that is,  there exists a $\sigmah{k}$-array $\DD=\{D_s\}_{s\in\Sigma^*}$ via a supporting function $f$ such that $\dom(f)$ is infinite, $\DD$ is honest, and $\DD$ intersects $A$. Since $f$ is componentwise honest, take an increasing polynomial $p$ such that $|x|\leq p(|y|)$ for any $x\in\dom(f)$ and any $y\in D_x$.

First, we define $S$ to be the collection of all nonempty forcing conditions $\sigma$ such that there exists an element $x\in\dom(f)$ satisfying that $D_{x}\subseteq\dom(\sigma)$ and $\sigma(y)=0$ for all $y\in D_{x}$. Note that $x\in\dom(f)$ implies $|x|\leq p(|\sigma|)$ since $\sigma$ is defined on all the strings in $D_x$. Let $q$ be any polynomial such that $|f(x)|\leq q(|x|)$ for all strings $x\in\dom(f)$. The set $S$ belongs to $\sigmah{k}$ because $S$ can be written as: 
\[
S=\{\sigma\mid \exists x\in\Sigma^{\leq p(|\sigma|)}\exists y\in\Sigma^{\leq q(|x|)} [(x,y)\in Graph(f)\wedge \forall i\in[1,m]_{\integer}( \sigma(y_i)\!\!\downarrow =0 )]\},
\]
where $y=\pair{y_1,y_2,\ldots,y_m}$. Next, we want to show that $S$ is dense. Let $\sigma$ be any forcing condition. Take any string $x$ such that $D_{x}\cap\dom(\sigma)=\setempty$ and $D_{x}\neq\setempty$. Such an $x$ exists because $\DD$ is eventually disjoint. For such an $x$, define $\tau$ as the unique forcing condition satisfying the following: $\sigma\subseteq\tau$,
$\dom(\tau)=\dom(\sigma)\cup D_{x}$, and $\tau(y)=0$ for all $y\in
D_{x}$. Clearly, $\tau$ is in $S$. This implies that $S$
is dense.

Since $A$ is $\sigmah{k}$-generic, we obtain $\sigma\subseteq A$ for a certain $\sigma$ in $S$. By the definition of $S$, there exists a string $x\in\dom(f)$ satisfying that $A(y)=0$ for all $y\in D_{x}$, which implies $D_x\cap A = \setempty$, a contradiction. Therefore, $A$ is honestly $\sigmah{k}$-hyperimmune.
\end{proofof}

In the late 1970s, Selman \cite{Sel79} introduced the notion of {\em $\p$-selective} sets, which are analogues of semi-recursive sets in recursion theory. These sets connect $\p$-immunity to $\p$-hyperimmunity. In general, for any class $\FF$ of {\em single-valued total} functions, we say that a set $S$ is {\em $\FF$-selective} if there exists a function (called the {\em selector}) $f$ in $\FF$ such that, for all pairs $(x,y)\in\Sigma^*\times\Sigma^*$, (i)
$f(x,y)\in\{x,y\}$ and (ii) $\{x,y\}\cap S\neq\setempty$ implies
$f(x,y) \in S$. For selectivity by multi-valued functions, the reader may refer to Hemaspaandra \etalc~\cite{HNOS96}.  Now, we consider the single-valued total function class $\sigmah{k}\mathrm{SV}_t$.

\begin{lemma}\label{lemma:selective}
Let $k\in\nat^{+}$. Every $\sigmah{k}$-immune 
$\sigmah{k}\mathrm{SV}_t$-selective set is honestly
$\sigmah{k}$-hyperimmune. 
\end{lemma}

Observe that the complement of a $\sigmah{k}\mathrm{SV}_t$-selective set $S$ is also $\sigmah{k}\mathrm{SV}_t$-selective because the exchange of the output string of any selector for $S$ gives rise to a selector for $\overline{S}$. Note also that Lemma \ref{lemma:selective} relativizes.

\begin{proofof}{Lemma \ref{lemma:selective}}
Let $k\geq1$ and assume that $S$ is $\sigmah{k}\mathrm{SV}_t$-selective but not honestly $\sigmah{k}$-hyperimmune. Our goal is to
show that $S$ has an infinite $\sigmah{k}$-subset. Let $f$ be any 
selector for $S$ and let $\DD = \{D_s\}_{s\in\Sigma^*}$ be any honest
$\sigmah{k}$-array intersecting $S$ via $g$. Define $h$ as follows. Let $y\in\dom(g)$ and assume that $D_{y}=\{x_{1},x_{2},\cdots,x_{m}\}$ with $x_1<x_2<\cdots<x_m$. Let $y_{1}=x_{1}$ and $y_{i+1}=f(y_{i},x_{i+1})$ for every $i\in[1,m-1]_{\integer}$ and then define $h(y)=y_{m}$. Clearly, $h$ is in $\sigmah{k}\mathrm{SV}$ since $Graph(h)$ is in $\sigmah{k}$.  For any string $y\in\dom(g)$, $h(y)$ belongs to $S$ since $D_y$ intersects $S$. Note that $h$ is honest because so is $\DD$. Let $p$ be any polynomial such that $|y|\leq p(|h(y)|)$ for any string $y$ in $\dom(h)$. Define $B=\{x\mid \exists
y\in\Sigma^{\leq p(|x|)}[y\in\dom(h) \:\wedge\: h(y)=x]\}$, which is in $\sigmah{k}$. Clearly, $B$ is a subset of $S$ and is infinite since $\DD$ is honest.
\end{proofof}

It follows from Lemma \ref{lemma:selective} that every $\np$-simple
$\p$-selective set is honestly $\np$-hypersimple since the complement
of any $\p$-selective set is also $\p$-selective.

Next, we show that strong $\p$-immunity does not imply honest
$\p$-hyperimmunity within the class $\e$. Earlier, Balc{\'a}zar
and Sch{\"o}ning \cite{BS85} created a strongly $\p$-bi-immune set $S$
in $\e$ with the density $|S\cap\Sigma^{\leq n}|=2^{n+1}-n-1$ for every number $n\in\nat$. For each $x\in\Sigma^*$, let $D_{x}$ consist of the first $|x|+2$
elements of $\Sigma^{|x|}$. Clearly, $D_{x}$ intersects $S$. This implies that $S$ is not honestly $\p$-hyperimmune. Therefore, we obtain the following proposition.

\begin{proposition}
There exists a strongly $\p$-bi-immune set in $\e$ that is 
not honestly $\p$-hyperimmune.
\end{proposition}

Next, we show $\p$-T-incompleteness of $\sigmah{k}$-hypersimple sets for each $k\geq1$.   

\begin{theorem}\label{theorem:hyper-closure}
Let $k\in\nat^{+}$.
\vs{-2}
\begin{enumerate}
\item No $\sigmah{k}$-hypersimple set is 
$\p$-T-complete for $\sigmah{k}$.
\vs{-2}
\item No honestly $\sigmah{k}$-hypersimple set is 
h-$\p$-T-complete for $\sigmah{k}$.
\end{enumerate}
\end{theorem}

Note that it is unclear whether we can replace $\p$-T-completeness in Theorem \ref{theorem:hyper-closure} by $\deltah{k}$-T-completeness. 
Now, we want to prove Theorem \ref{theorem:hyper-closure}. Our proof utilizes the lemma below.

\begin{lemma}\label{lemma:hyper-almost}
Let $k$ be any number in $\nat^{+}$ and let $A$ be any infinite set in $\sigmah{k}$.
\begin{enumerate}\vs{-2}
\item If $A\Treduces{\p}B$ and $B$ is $\sigmah{k}$-hyperimmune and in $\dexp$, then $\overline{A}$ is almost $\deltah{k}$-immune.
\vs{-2}
\item If $A\Treduces{h\mbox{-}\p}B$ and $B$ is honestly $\sigmah{k}$-hyperimmune, then $\overline{A}$ is almost $\deltah{k}$-immune.
\end{enumerate}
\end{lemma}

We postpone the proof of Lemma \ref{lemma:hyper-almost} and instead prove Theorem \ref{theorem:hyper-closure}.

\begin{proofof}{Theorem \ref{theorem:hyper-closure}}
We prove only the first claim since the second claim follows similarly. Assume that $B$ is a $\sigmah{k}$-hypersimple set that is $\p$-T-complete for $\sigmah{k}$. This means that $\overline{B}$ is $\sigmah{k}$-hyperimmune and is in $\pih{k}$. Clearly, the existence of a $\sigmah{k}$-hypersimple set implies $\deltah{k}\neq\sigmah{k}$. Note that $B$ is in $\dexp$. Since $B$ is $\p$-T-complete for $\sigmah{k}$, every $\sigmah{k}$-set is $\p$-T-reducible to $B$. {}From Lemma \ref{lemma:hyper-almost}, it follows that every $\pih{k}$-set is almost $\deltah{k}$-immune. This contradicts Lemma \ref{lemma:paddable}, in which $\pih{k}$ has a $\deltah{k}$-levelable set. Therefore, $B$ cannot be $\sigmah{k}$-hypersimple.
\end{proofof}

We still need to prove Lemma \ref{lemma:hyper-almost}, which requires a key idea of Agrawal (mentioned earlier), who showed that no $\np$-simple set is h-$\p$-btt-complete for $\np$. We extend his core argument to Lemma~\ref{key-lemma}. For convenience, we say that a complexity class $\CC$ is {\em closed under intersection with $\deltah{k}$-sets} if, for any set $A$ in $\CC$ and any set $B$ in $\deltah{k}$, the intersection $A\cap B$ belongs to $\CC$.

\begin{lemma}\label{key-lemma}
Let $\CC$ be any complexity class containing $\deltah{k}$ such that $\CC$ is closed under intersection with $\deltah{k}$-sets. Let $A$ be any set in $\CC$ whose complement is $\deltah{k}$-levelable. If $A$ is $\deltah{k}$-T-reducible to $B$ via a reduction machine $M$, then there exists an infinite set $C$ in $\CC$ such that $Q(M,B,x)\cap B\neq\setempty$ for all $x\in C$.
\end{lemma}

\begin{proof}
Assume that $A$ is $\deltah{k}$-T-reducible to $B$ via a certain reduction machine $M$; namely, $A=\{x\mid M^{B}(x)=1\}$. For convenience, introduce the set $E=\{x\mid M^{\setempty}(x)=0\}$, which is obviously in $\deltah{k}$.

First, consider the case where $|E\cap A|$ is infinite.
In this case, for every $x\in E\cap A$, $M^{B}(x)=1$ but $M^{\setempty}(x)=0$. Hence, $M$ on input $x$ must query a certain string in $B$,  which implies $Q(M,B,x)\cap B\neq\setempty$. Let $C=E\cap A$. Clearly, $C$ is infinite and is in $\CC$ because $A$ is in $\CC$ and $\CC$ is closed under intersection with $\deltah{k}$-sets.

Second, we consider the other case where $|E\cap A|$ is finite.  Let
$E'=E \setminus A$. Since $E'$ differs from $E$ on finitely many  elements, $E'$ belongs to $\deltah{k}$. Note that $E'\subseteq\overline{A}$ by its
definition. Using the assumption that $\overline{A}$ is $\deltah{k}$-levelable, there exists an infinite set $C\in \deltah{k}$ such that $C\subseteq \overline{A}$ and $C\cap E'=\setempty$. Let $x$ be any string in $C$. Since $x\in \overline{A}$, $M^{B}(x)$ outputs $0$. Nonetheless, from $x\not\in E$, $M^{\setempty}(x)$ equals $1$. Thus, $Q(M,B,x)\cap B\neq\setempty$. Obviously, $C$ is in $\CC$ (because $\deltah{k}\subseteq\CC$) and is clearly infinite.
\end{proof}

Now, we give the proof of Lemma \ref{lemma:hyper-almost}.

\begin{proofof}{Lemma \ref{lemma:hyper-almost}}
2) Let $A$ be any infinite $\sigmah{k}$-set and let $B$ be any  honestly $\sigmah{k}$-hyperimmune set. Assume that $A$ is h-$\p$-T-reducible to $B$ via a certain reduction machine $M$.  Our goal is to show that $\overline{A}$ is almost $\deltah{k}$-immune. Assume to the contrary that $\overline{A}$ is a $\deltah{k}$-levelable set in $\pih{k}$. By Lemma \ref{key-lemma}, there exists an infinite set $C$ in $\sigmah{k}$ such
that $Q(M,B,x)\cap B\neq\setempty$ for all $x\in C$. 

\begin{claim}
For any string $x$, $Q(M,B,x)\cap B \neq \setempty$ if and only if $Q(M,\setempty,x)\cap B \neq \setempty$.
\end{claim}

The proof of the above claim proceeds as follows. Assume that $Q(M,B,x)\cap B=\setempty$. This means that all queries of $M$ on input $x$ with oracle $B$ are answered NO. Thus, $Q(M,B,x)=Q(M,\setempty,x)$, which implies $Q(M,\setempty,x)\cap B=\setempty$. The other direction is similar.

We return to the main argument. The partial function $h$ is defined as a map from $\Sigma^*$ to $\Sigma^*$ as follows. The set $C$ is the domain of $h$. Choose any string $x$ in $C$ and consider the set $Q(M,\setempty,x)$. Note that $|Q(M,\setempty,x)|$ is polynomially bounded. Moreover, $Q(M,\setempty,x)$ cannot be empty by the above claim. Let $h(x)$ be the unique string that represents $Q(M,\setempty,x)$. Now, define $D_x$ to be $set(h(x))$ if $x\in C$, and $D_x=\setempty$ otherwise. 

The sequence $\DD = \{D_x\}_{x\in\Sigma^*}$ satisfies $D_x\cap B\neq\setempty$ for all $x\in\dom(h)$. Hence, $\DD$ is a honest $\sigmah{k}$-array because $M$ is honest and $C$ is in $\sigmah{k}$. This contradicts our assumption that $B$ is honestly $\sigmah{k}$-hyperimmune.

1) Unlike 2), we now need to prove that the array $\DD = \{D_x\}_{x\in\Sigma^*}$ defined in 2) is eventually disjoint. First, assume that $M$ runs within time $n^d+d$ for all inputs of length $n$ and any oracle, where $d$ is a fixed positive constant. Unfortunately, since $M$ is not guaranteed to be honest, we cannot prove the eventual-disjointness of $\DD$. To overcome this problem, we modify the reduction machine $M$ as follows.  

Since $B$ is in $\dexp$, there exists a exponential-time deterministic TM $N$ that recognizes $B$. Without losing generality, we can assume that $M$ runs within time $2^{n^d+d}$ for any input of length $n$. We modify the reduction machine $M$ as follows: 
\begin{quote}
{\sf On input $x$, simulate $M$ on input $x$. If $M$ makes a query $y$, then  check whether $|y|^d > \log{|x|}$. If so, make a query $y$ to oracle. Otherwise, run $N$ on input $x$ and make its outcome an oracle answer.} 
\end{quote}
Let $M_{+}$ be the new oracle TM defined above. Note that $M_{+}$ is an oracle $\p$-machine because its running time is, for a certain fixed constant $c>0$, bounded above by $c(|x|^d+d+2^{\log{|x|}+d})$, which is $O(|x|^d)$. It is thus clear that $M_{+}$ $\p$-T-reduces $A$ to $B$. 
Note that $M_{+}$ on input $x$ cannot query any string shorter than length $\log{|x|}$. Therefore, $M_{+}$ satisfies the following condition: for every $x$, there exists a string $y$ with $y\geq x$ such that (i) $Q(M_{+},B,x)\cap Q(M_{+},B,y)=\setempty$ and (ii) if $Q(M_{+},B,y)\neq\setempty$, then $Q(M_{+},B,y)$ surpasses $Q(M_{+},B,x)$. 

Now, consider the array $\{D_x\}_{x\in\Sigma^*}$ obtained from $M_{+}$ similar to 2). Since $D_x=Q(M_{+},\setempty,x)$ for all $x\in\dom(h)$, it immediately follows that $\DD$ is eventually disjoint. 
\end{proofof}

Although the existence of a $\sigmah{k}$-simple set is unknown,
as Schaefer and Fenner \cite{SF98} demonstrated, it is relatively easy to prove the existence of an honest $\np^G$-hypersimple set relative to a Cohen-Feferman generic oracle $G$. For a higher level $k$ of the polynomial hierarchy, we build in the following proposition a recursive oracle relative to which an honest  $\sigmah{k}$-hypersimple set exists.

\begin{proposition}\label{exist-hypersimple}
For each $k\in\nat^{+}$, there exists a recursive oracle $A$ such that an honest $\sigmah{k}(A)$-hypersimple set exists.
\end{proposition}

We prove Proposition \ref{exist-hypersimple} using Bruschi's \cite{Bru92} result and Lemma \ref{lemma:selective}. To use Lemma \ref{lemma:selective}, we need the following supplemental lemma. Recall the tower set $\hat{T}$ introduced in Section \ref{sec:strong-immune}.

\begin{lemma}\label{tower-selective}
Let $A$ be any set in $\dexp$. If $A\subseteq\{1^n\mid n\in \hat{T}\}$, then $A$ is $\p$-selective.
\end{lemma}

\begin{proof}
Since $A\in\dexp$, there exists a number $c\in\nat$ and a deterministic TM $M$ that recognizes $A$ in at most $2^{n^c+c}$ steps, where $n$ is the length of an input. Consider the following function $f$. On input $(x,y)\in\Sigma^*\times\Sigma^*$, if $|x|\not\in T$ and $|y|\not\in T$, then $f$ outputs the lexicographically minimal string between $x$ and $y$. Otherwise, there are three possibilities: $|y|\leq \log\log|x|$, $|x|\leq \log\log|y|$, or $x=y$. If $x=y$, then output $x$. Assume that $|y|\leq \log\log|x|$. In this case, $f$ outputs $y$ if $M(y)=1$, and $f$ outputs $x$ otherwise. Note that the running time of $M$ on input $y$ is bounded above by $2^{|y|^c+c}$, which is at most $|x|^d+d$ for a certain constant $d\geq1$ depending only on $c$. For the last case, $f$ outputs $x$ if $M(x)=1$ and $y$ otherwise. Obviously, $f$ is deterministically computed in polynomial time and thus, $f$ belongs to $\fp$. It is easy to verify that $f$ satisfies the selectivity condition.
\end{proof}

Note that Lemma \ref{tower-selective} relativizes. Using this lemma, we can prove Proposition \ref{exist-hypersimple}.

\begin{proofof}{Proposition \ref{exist-hypersimple}}
This proof is similar to that of Proposition \ref{strong-oracle}. For each $k\in\nat^{+}$, we first recall the oracle-dependent set $L^A_k$ defined in the proof of Proposition \ref{strong-oracle}. Note that, for any oracle $A$, $L_k^A$ belongs to $\pih{k}(A)$ and thus to $\dexp^A$. Applying Lemma \ref{tower-selective} to $L^A_k$, we obtain the $\p^A$-selectivity of $L^A_k$. As noted in the proof of Proposition \ref{strong-oracle}, we can build a recursive oracle $A$ such that $L_k^A$ is $\sigmah{k}(A)$-immune. {}From Lemma \ref{lemma:selective}, it immediately follows that $L_k^A$ is honestly $\sigmah{k}(A)$-hyperimmune.
\end{proofof}

We also show a random oracle result on the existence of honestly $\np$-hypersimple set using Lemmas \ref{lemma:selective} and \ref{tower-selective}. 

\begin{proposition}\label{prop:hyper-random}
An honestly $\np^X$-hypersimple set exists
relative to a random oracle $X$.
\end{proposition}

\begin{proof}
Recall from the proof of Proposition \ref{strong-random} that the set 
$K^A$ is in $\co\np^A$ for any oracle $A$. By Lemma \ref{tower-selective}, $K^A$ is $\p^A$-selective for any oracle $A$. We already noted in the proof of Proposition \ref{strong-random} that $K^X$ is $\np^X$-immune relative to a random oracle $X$. Since Lemma \ref{lemma:selective} relativizes, $K^X$ is honestly $\np^X$-hyperimmune relative to a random oracle $X$.
\end{proof}

An important open problem is to prove that, at each level $k$ of the polynomial hierarchy, an honest $\sigmah{k}$-hypersimple set exists relative to a random oracle.

\section{Limited Immunity and Simplicity}
\label{sec:k-immune}

By current techniques, we cannot determine whether an $\np$-simple set exists. The difficulty comes from the fact
that an $\np$-immune set requires {\em every} $\np$-subset to be finite. If we restrict our attention to certain types of $\np$-subsets, then we may overcome this difficulty. Under the name of $k$-immune sets, Homer \cite{Hom86} required only $\np_{(k)}$-subsets, for a fixed number $k$, to be finite. He then demonstrated how to construct a $k$-simple set within $\np$. In this section, we investigate the notions obtained by restricting the conditions of immunity and simplicity. We first review Homer's notions of $k$-immunity and $k$-simplicity in Definition \ref{def:k-immune}.
Recall from Section \ref{sec:notation} the identification between $\Sigma^*$ and $\nat$. Here, we freely identify binary strings with natural numbers. 

\begin{definition}\label{def:k-immune}{\rm \cite{Hom86}}\hs{2}
Let $k\in\nat^{+}$. 
\vs{-2}
\begin{enumerate}
\item A set $S$ is {\em $k$-immune}  
if $S$ is infinite and there is no index $i$ in $\gindex_{(k)}$ such that $W_{i}\subseteq S$ and $W_{i}$ is infinite. 
\vs{-2}
\item A set $S$ is {\em $k$-simple} 
if $S$ is in $\np$ and $\overline{S}$ is $k$-immune. 
\end{enumerate}
\end{definition}

Obviously, any $k$-immune set is $k'$-immune for any $k'\leq k$ since $\gindex_{(k')}\subseteq \gindex_{(k)}$. Similarly, any $k$-simple set is $k'$-simple if $k'\leq k$.

Homer \cite{Hom86} constructed a $k$-simple set for each
$k\in\nat^{+}$ using Ladner's \cite{Lad75} delayed diagonalization technique. His $k$-simple set $A$ satisfies the following {\lq\lq}sparseness{\rq\rq} property: for each number $n\in\nat$, the cardinality $|A\cap\Sigma^n|$ is at most $2\log{n}$. By analyzing his construction, we can prove the following lemma, which will play a key role in the proof of Theorem \ref{lemma:not_feasibly_ksimple}. 

\begin{lemma}\label{lemma:not_feasibly_ksimple3}
For every integer $k \geq 1$, there exists a $k$-simple set $A$ with the following property: $|A\cap\Sigma^n|$ is at most $2\log \log n$ for each number $n\in\nat$.
\end{lemma}

An {\lq\lq}effective{\rq\rq} version of immune and simple sets, called {\em effectively immune} and {\em effectively simple sets}, has been studied in recursion theory for the class $\mathrm{RE}$. Effectively simple sets are known to be T-complete for $\mathrm{RE}$ and there also exists an effectively simple tt-complete set for $\mathrm{RE}$. If $A$ is strongly effectively immune, then $\overline{A}$ cannot be immune. The reader may refer to \eg Odifreddi's \cite{Odi89} textbook for more details.  Recently, Ho and Stephan \cite{HS02} constructed a simple set to which any effectively simple set can be 1-reducible. Analogously, we consider a resource-bounded version of such effectively immune sets and effectively simple sets.

\begin{definition}\label{def:feasible-immune}
Let $k\in\nat^{+}$. 
\vs{-2}
\begin{enumerate}
\item A set $S$ is {\em feasibly $k$-immune} if (i) $S$ is infinite and
(ii) there exists a polynomial $p$ such that, for all indices $i\in
\gindex_{(k)}$, $W_{i}\subseteq S$ implies $|W_{i}|\leq2^{p(i)}$.
\vs{-2}
\item A set $S$ is {\em feasibly $k$-simple} if $S\in\np$ and
$\overline{S}$ is feasibly $k$-immune.
\end{enumerate}
\end{definition}

Obviously, every feasibly $k$-immune set is $k$-immune for any number $k\in\nat^{+}$. Using a straightforward diagonalization, we can construct a feasibly $k$-immune set that falls in $\deltah{2}$. 

\begin{proposition}\label{exist-feasible}
Let $k\in\nat^{+}$. There exists a feasibly $k$-immune set in $\deltah{2}$.
\end{proposition}

The desired set $A$ that we will construct in the following proof is not honestly $\p$-hyperimmune. This comes from the fact that $A$ satisfies the property $|\overline{A}\cap\Sigma^n|\leq n$ for all numbers $n\in\nat^{+}$. For each $x$, the set $D_x$ of the first $|x|+1$ elements in $\Sigma^{|x|}$ clearly intersects $A$.

\begin{proofof}{Proposition \ref{exist-feasible}}
First, we fix $k$ arbitrarily. For each $i,s\in\nat$, let $\varphi_{i,s}$ denote the machine obtained from $\varphi_i$ by restricting its running time as follows: let $\varphi_{i,s}(x)=\varphi_i(x)$ if $\varphi_i(x)$ terminates within step $s$ and let $\varphi_{i,s}(x)$ be undefined otherwise. We wish to construct the desired feasibly $k$-immune set $A=\bigcup_{i\in\nat}A_i$ by stages, where each $A_i$ is a subset of $\Sigma^{\leq n}$. 

During our construction process, we intend to meet the following two requirements:
\begin{itemize}
\item[1)] $R^{(0)}_i$: $|\overline{A}\cap \Sigma^i|\leq i$.
\vs{-2}
\item[2)] $R^{(1)}_i$: if $|W_i|>2^{i+1}$ and $i\in \gindex_{(k)}$, then $W_i\cap \overline{A} \neq \setempty$.
\end{itemize}
The first requirement makes $A$ infinite. The second requirement implies that, for any index $i$ in $\gindex_{(k)}$, if $W_i\subseteq A$ then $|W_i|\leq 2^{i+1}$. {}From these two requirements, we therefore obtain the feasible $k$-immunity of $A$.
We also build the set $\marked_i$ to mark all used indices at stage $i$ and finally let $\marked = \bigcup_{i\in\nat}\marked_i$. For any finite subset $B$ of $\Sigma^*$, the notation $\max(B)$ simply denotes the lexicographically maximal string in $B$. Since $B\subseteq \Sigma^{\leq |\max(B)|}$, it holds that $|B|\leq |\Sigma^{\leq |\max(B)|}| \leq 2^{|\max(B)|+1}$. For brevity, write $\ell(i,j,k)$ for $|i|\cdot|j|^k+|i|$. Our construction proceeds as follows:
\begin{quote}
{\sf Stage 0:} Let $A_0=\setempty$ and $\marked_0=\setempty$.

{\sf Stage $n\geq1$:} At this stage, we consider all the strings of length $n$. Consider the following set:
\[
 C=\{\pair{i,j}\mid i\leq |j| \:\wedge\: \varphi_{i,\ell(i,j,k)}(j)\!\!\downarrow=1 \:\wedge\: |\{x < j\mid \varphi_{i,\ell(i,x,k)}(x)\!\!\downarrow=1\}|> i\}.
\]
For each $i\leq n$, let $j_{n,i}$ be the lexicographically maximal string $j$ in $\Sigma^n$ such that $\pair{i,j}\in C$  if $j$ exists. Otherwise, let $j_{n,i}$ be undefined. Consider the set $C_n$ of all  indices $i\leq n$ for which $j_{n,i}$ exists. Define $\marked_n$ to be $\marked_{n-1}\cup C_n$. Finally, define $A_n$ to be $A_{n-1}\cup (\Sigma^n\setminus \{j_{n,i}\mid i\in C_n\})$. 
\end{quote}

First, we show that $C$ belongs to $\np$. Obviously, we need an $\np$-computation to determine whether $\varphi_{i,\ell(i,j,k)}(j)\!\!\downarrow=1$. Moreover, we can determine whether the cardinality $|\{x < j\mid \varphi_{i,\ell(i,x,k)}(x)\!\!\downarrow=1\}|$ is more than $i$ by nondeterministically guessing $i+1$ $x$'s that satisfy $\varphi_{i,\ell(i,x,k)}(x)\!\!\downarrow=1$.  Since $i+1\leq |j|+1$, this process requires only an $\np$-computation. Therefore, $C$ belongs to $\np$.
To compute each set $A_i$, we need to compute all the elements in $C_i$, \ie all the well-defined $j_{n,i}$'s. This is done by a standard binary search technique using $C$ as an oracle in polynomial time. Therefore, $A$ is in $\p^{C}\subseteq \p^{\np}$. 

It remains to show that the two requirements $R^{(0)}_i$ and $R^{(1)}_i$ are satisfied at every stage $i$.

\begin{claim}
$R^{(0)}_i$ is satisfied at each stage $i$.
\end{claim}

Note that $|C_i|\leq i$ at each stage $i$. Hence, we obtain $|\overline{A}\cap \Sigma^i|\leq |C_i| \leq i$, which clearly meets  the requirement $R^{(0)}_i$.

\begin{claim}
$R^{(1)}_i$ is satisfied for any number $i$.
\end{claim}

To prove this claim, it suffices to show that, for any index $i$ in $\gindex_{(k)}$, if $|W_i|>2^{i+1}$ then $i$ is in $\marked$ because $i\in \marked$ means that, at a certain stage $n$, there exists the unique string $j_{n,i}$ in $\Sigma^n$ such that $j_{n,i}\in W_i$ and $j_{n,i}\in \overline{A}$. Now, suppose that $i\in \gindex_{(k)}$ and $|W_i|>2^{i+1}$. Recall that $W_i=\{x\mid \varphi_{i,\ell(i,x,k)}(x)\!\!\downarrow=1\}$. 

{\sf (Case 1)} Consider the case where $W_i$ is finite and $i>|\max(W_i)|$. This case never happens because $|W_i|\leq  2^{|\max(W_i)|+1}<2^{i+1}$. 

{\sf (Case 2)} Assume that either $W_i$ is infinite or $i\leq |\max(W_i)|$. We split this case into two subcases.

{\sf (Case 2a)} Consider the case where $W_i$ is finite and $i\leq |\max(W_i)|$. Define $j_0$ as $\max(W_i)$ and let $n=|j_0|$. We  want to show that $i\in \marked$. Assume otherwise that $i\not\in \marked$. Since $j_0\in W_i$, we have $\varphi_{i,\ell(i,j_0,k)}(j_0)\!\!\downarrow=1$. Clearly, $i\leq |j_0|$. Recall that $C_n=\{i\mid i\leq n \:\:\wedge\:\: j_{n,i} \text{ exists}\}$. Since $i\not\in \marked$, $i$ cannot belong to $C_n$. This means that $j_{n,i}$ does not exist, and thus we obtain $\pair{i,j}\not\in C$ for all $j\in\Sigma^{n}$. In particular, $\pair{i,j_0}\not\in C$ because $j_0\in\Sigma^n$. By the definition of $C$, we can conclude that $|\{x\in W_i\mid x<j_0\}|\leq i$.  
It therefore follows from $j_0=\max(W_i)$ that $|W_i|=|\{x\in W_i\mid x<j_0\}|+1\leq i+1$, which implies $|W_i|\leq 2^{i+1}$, a contradiction. Hence, $i$ is in $\marked$.

{\sf (Case 2b)} If $W_i$ is infinite, then let $j_1$ be the lexicographically $2^{i+1}+1$st string in $W_i$. Assume that $i\not\in \marked$. Similar to Case 2a, this leads to the conclusion that $|\{x\in W_i\mid x<j_1\}|\leq i$. However, the choice of $j_1$ implies $|\{x\in W_i\mid x<j_1\}|=2^{i+1}>i$, a contradiction. Therefore, $i$ is in $\marked$, as requested. 
\end{proofof}

{}From Definition \ref{def:feasible-immune}, it follows that every feasibly $k$-immune set is $k$-immune. The converse, however, does not hold since there exists a $k$-simple set which is not feasibly $k$-simple for each number $k\in\nat^{+}$. The theorem below is slightly stronger than this claim because any feasibly $k$-simple set is feasibly $1$-simple. 

\begin{theorem} \label{lemma:not_feasibly_ksimple}
For each $k\in\nat^{+}$, there exists a $k$-simple set which is not feasibly 1-simple.
\end{theorem}

Hereafter, we prove Theorem \ref{lemma:not_feasibly_ksimple}. To show this, we use the language $L_u = \{ uv \mid |uv|=2^{2^{|u|}}\}$, where $u$ is any fixed string. Observe that, for each string $u$, there exists a G{\"o}del number $i$ of length $O(|u|)$ for which $\varphi_i$ recognizes $L_u$. More precisely, the following lemma holds. 

\begin{lemma}\label{lemma:not_feasibly_ksimple2}
There exists a number $a$ in $\nat$ that satisfies the following condition: for every string $u$ with $|u|\geq a$, there exists an index $i$ such that (i) $W_i = L_u$, (ii) $|u| < |i| < a(|u| + 1)$, and (iii) the running time of ${\varphi}_i$ on input $x$ is at most $|i|(|x|+1)$ for all $x\in W_i$.
\end{lemma}

Lemma \ref{lemma:not_feasibly_ksimple2} can be shown by directly constructing an appropriate deterministic TM that recognizes $L_u$.  We leave its proof to the reader. 
Now, we are ready to present the proof of
Theorem~\ref{lemma:not_feasibly_ksimple}.

\begin{proofof}{Theorem~\ref{lemma:not_feasibly_ksimple}}
Let $k\geq 1$ and let $A$ be any $k$-simple set that satisfies
Lemma~\ref{lemma:not_feasibly_ksimple3}. We want to show that $A$ is
not feasibly $1$-simple. Let $L_u=\{uv\mid |uv|=2^{2^{|u|}}\}$ for each string $u\in\Sigma^*$. 

Take any natural number $a$ for which
Lemma~\ref{lemma:not_feasibly_ksimple2} holds. 
Consider the following claim. 

\begin{claim}
For each number $\ell\in\nat$, there exists an index $i$ such that the following four conditions hold: (i) $W_i \subseteq \overline{A}$, (ii) $\ell < |i|$, (iii)  $\log \log {|W_i|} \geq (i^{1/a})/2 -1$, and (iv) the running time of ${\varphi}_i$ on
input $x$ is at most $|i|(|x|+1)$ for any string $x$ in $W_i$.
\end{claim}

Assume for a while that the above claim holds. The claim guarantees that, for every polynomial $p$, there exists an index $i\in \gindex_{(1)}$ satisfying that $W_i\subseteq \overline{A}$ and $|W_i|>2^{p(i)}$. Therefore, $\overline{A}$ is not feasibly
$1$-immune and this will complete the proof. 

Now, we prove the claim. Let $\ell$ and $m$ be
any natural numbers satisfying that $m \geq \max \{ a, \ell \}$ and
$2m < 2^m$. Define $n=2^{2^m}$, which is equivalent to $m=\log\log{n}$.  The condition $|A \cap \{ 0,1
\}^n|\leq 2m$ implies that the cardinality of the set $\{ uv
\mid |uv|=n \:\wedge\: |u|=m\}$ is at most $2m$. The set $\{ u\in \Sigma^m \mid L_u \cap A \ne \setempty \}$ therefore has the cardinality at most $2m$.  Since $2m<2^m$, we have at least one string $u$ of length $m$ satisfying that $L_u \subseteq \overline{A}$. Fix such a string $u$. Clearly, $|L_u|=2^{n-m}\geq 2^{n/2}$ since $2m<n$.
By Lemma~\ref{lemma:not_feasibly_ksimple2}, a certain index $i$
satisfies the following three conditions: $W_i = L_u$, $m < |i| < a(m + 1)$, and the running time of ${\varphi}_i$ on input $x$ is at most $|i|(|x|+1)$ for any $x\in W_i$. For such an $i$, it follows that
$(2\log n)^a = 2^{a(m + 1)} \geq 2^{|i|+1} \geq i$, which implies $\log n \geq \frac{1}{2}i^{1/a}$. 
Therefore, $\log \log {|W_i|} 
\geq \log (n/2)$, which is at least $\frac{i^{1/a}}{2} -1$. 
\end{proofof}

We return to the old question of whether $\np$-simple sets exist.
Unfortunately, there seems no strong evidence that suggests the existence of such a set. Only relativization provides a world where $\np$-simple sets exist. At the same time, there is another relativized world where these sets do not exist. These relativization results clearly indicate that the question of whether $\np$-simple sets exist needs nonrelativizable proof techniques.

In the past few decades, the {\em Berman-Hartmanis isomorphism conjecture} \cite{BH77} has served as a working hypothesis in connection to $\p$-m-complete sets for $\np$. By clear contrast, there has been no {\lq\lq}natural{\rq\rq} working hypothesis that yields the existence of $\np$-simple sets. For example, the hypothesis $\p\neq\np$ does not suffice since Homer and Maass \cite{HM83} constructed a relativized world where the assumption $\p\neq\np$ does not imply the existence of an $\np$-simple set. Motivated by Homer's $k$-simplicity result, we propose the following working hypothesis:

\begin{quote}
{\sf The k-immune hypothesis:} There exists a positive integer $k$
such that every infinite $\np$ set has an infinite $\np_{(k)}$-subset.
\end{quote}

\n Under this hypothesis, we can derive the desired consequence: the existence of $\np$-simple sets.

\begin{lemma}
If the k-immune hypothesis holds, then there exists an $\np$-simple set.
\end{lemma}

\begin{proof}
Assume that the k-immune hypothesis is true. There exists a positive integer $k$ such that every infinite $\np$-set has an infinite $\np_{(k)}$-subset. Consider any $k$-simple set $A$. We claim that $A$ is $\np$-simple. If $A$ is not $\np$-simple, then $\overline{A}$ has an infinite $\np$-subset $B$. By our hypothesis, $B$ contains an infinite $\np_{(k)}$-subset. Hence, $A$ cannot be $k$-simple, a contradiction. Therefore, $A$ is $\np$-simple.
\end{proof}

To close this section, we claim that the k-immune hypothesis fails relative to a Cohen-Feferman generic oracle. 

\begin{proposition}\label{prop:k-immune}
The $k$-immune hypothesis fails relative to a Cohen-Feferman generic
oracle.
\end{proposition}

In the following proof, we use {\em weak forcing} as in the proof of Theorem \ref{theorem:strongly-simple}. We also assume that, for each  number $k\in\nat^{+}$, there exists a list
$\langle N_{(k),i} \mid i \in \nat \rangle$ 
of nondeterministic oracle TMs such that ${\np}_{(k)}^A =
\{ L(N_{(k),i},A) \mid i \in \nat \}$ for any oracle $A$, and each
$N_{(k),i}$ is clocked by $|i|\cdot n^k + |i|$, where $n$ is the length of an input. In the following proof, we fix such a list. 

\begin{proofof}{Proposition \ref{prop:k-immune}} 
Let $k$ be any number in $\nat^{+}$. Letting $G$ be any Cohen-Feferman generic set, we show the existence of an infinite set in ${\p}^G$ that has no infinite ${\np}_{(k)}^G$-subset. Consider the set $L_k^X = \{ x \mid x0^{|x|^{k+2}} \in X \}$ for each oracle $X$. Clearly, $L_k^G$ belongs to ${\p}^G$. For each fixed number $n\in\nat$, let ${\cal D}^n_1$ denote the set of all forcing conditions $\sigma$ that force {\lq\lq}$\exists x \in \Sigma^*\: [x\in \Sigma^{< n} \:\vee\: x0^{|x|^{k+2}}\in X ]$,{\rq\rq} where $X$ is a variable running over all subsets of $\Sigma^*$. Since  ${\cal D}^n_1$ is obviously dense for every $n$, $L_k^G$ is indeed an infinite set.

Now, let $S$ be any ${\np}_{(k)}^G$-subset of $L_k^G$. We need to show that $S$ is finite. Since $S \in {\np}_{(k)}^G$, there exists an index $i\in\nat$ for which $S = L(N_{(k),i},G)$. Take any sufficiently large natural number $c$ such that $|i| \cdot |x|^k + |i| < |x|^{k+2}$ for every $x\in\Sigma^*$ with $|x| \geq \max \{ |i|,c \}$. Hence, the machine $N_{(k),i}$ does not query any string of length $\geq |x|^{k+2}$ on input $x$ with any oracle. Consider the following two statements:

\begin{itemize}
\item $\varphi_0(X) \equiv$ $\exists\, n\in\omega\: \forall\, x\in \Sigma^*\: [x\not\in L(N_{(k),i},X) \:\vee\: |x|\leq n]$, and
\vs{-2}
\item $\varphi_1(X) \equiv$ $\exists\, x \in \Sigma^*\: [x\in L(N_{(k),i}, X) \:\wedge\: x0^{|x|^{k+2}} \not\in X ]$.
\end{itemize}

\n Let ${\cal D}_2$ be the set of all forcing conditions $\sigma$ that force $(\varphi_0(X) \vee \varphi_1(X))$. We want to claim that $(\varphi_0(G)\:\vee\:\varphi_1(G))$ is true. To prove this claim, it suffices to show that $\DD_2$ is dense.

Now, we prove that $\DD_2$ is dense. If there is an extension $\tau$ of $\sigma$ that forces $\varphi_0(X)$, then $\tau$ is in $\DD_2$ and also forces $(\varphi_0(X) \vee \varphi_1(X))$. Next, assume that no extension of $\sigma$ forces $\varphi_0(X)$. By the property of weak forcing, this implies that $\sigma$ forces $\neg\varphi_0(X)$, which means {\lq\lq}$\forall\: n\in\omega\: \exists\: x\in \Sigma^* \: [ N_{(k),i}^X(x)=1 \:\wedge\: |x|> n]$.{\rq\rq} Choose any sufficiently large number $n_0$ such that $\sigma$ forces {\lq\lq}$\exists x\in \Sigma^*\: [x\in L(N_{(k),i},X) \:\wedge\: |x|> n_0]$.{\rq\rq} Furthermore, choose a string $x$ and an extension $\rho$ of $\sigma$ satisfying that $\rho$ forces {\lq\lq}$N_{(k),i}^X(x)=1 \:\wedge\: |x|> n_0$.{\rq\rq} Since $N^{\rho}$ on input $x$ cannot query any string of length $\geq |x|^{k+2}$, we can assume that $\dom(\rho)$ does not include $x0^{|x|^{k+2}}$. Therefore, we can define the extension $\tau$ of $\rho$ so that $\dom(\tau)=\dom(\rho)\cup\{x0^{|x|^{k+2}}\}$ and $\tau(x0^{|x|^{k+2}})=0$. Clearly, $\tau$ forces $\varphi_1(X)$ and thus, it forces  $(\varphi_0(X) \vee \varphi_1(X))$. Therefore, $\DD_2$ is dense.

Since $G$ is generic, either $L(N_{(k),i}, G)$ is finite or $L(N_{(k),i},G) - L_k^G$ is nonempty. Recall that $S=L(N_{(k),i},G)$. Since $S$ is a subset of $L_k^G$, $S$ is therefore finite, as required.
\end{proofof} 

A remaining open problem is to construct a relativized world where the k-immune hypothesis holds.

\section{Completeness under Non-Honest Reductions}
\label{sec:complete}

Immunity has a deep connection to various completeness notions since Post's attempt to build a non-recursive Turing-incomplete 
{r.e.\!\!} set. For
example, there is a simple, tt-complete set for $\mathrm{RE}$; however, no simple set is btt-complete for $\mathrm{RE}$; see, \eg Odifreddi's \cite{Odi89} textbook for more results. In the previous sections, we have shown that various types of resource-bounded simple sets cannot be complete under certain polynomial-time honest reductions.  This section instead studies incompleteness of simple sets under non-honest reductions.

To remove the honesty condition from reductions, we often need to make extra assumptions for similar incompleteness results. In mid 1980s, Hartmanis, Li, and Yesha \cite{HLY86} proved that (i) no $\np$-immune set in $\dexp$ is $\p$-m-hard for $\np$ if $\np\nsubseteq\subexp$ and (ii) no $\np$-simple set is $\p$-m-complete if $\np\cap\co\np\nsubseteq\subexp$. These results can be expanded to any $\Delta$-level of the polynomial hierarchy and of the subexponential hierarchy. We also improve the latter result, which follows from our main theorem on the $\deltah{k}$-1tt-completeness.

\begin{proposition}\label{prop:p-m-hard} 
Let $j$ and $k$ be any nonnegative integers. 
\begin{enumerate}\vs{-2}
\item No $\sigmah{k}$-immune set in $\deltaexph{j}$ is $\deltah{k}$-m-hard for $\sigmah{k}$ if $\sigmah{k} \nsubseteq\subdeltaexph{\max \{ j,k \}}$.
\vs{-2}
\item No $\sigmah{k}$-simple set is $\deltah{k}$-m-complete for $\sigmah{k}$ if $\unique(\sigmah{k}\cap\pih{k}) \nsubseteq\subdeltaexph{k}$.
\end{enumerate}
\end{proposition}

Note that Proposition~\ref{prop:p-m-hard}(1) follows from Lemma
\ref{lemma:hartmanis}(1) and Proposition \ref{prop:p-m-hard}(2) is a direct consequence of Theorem \ref{prop:1tt-hard}(2) because every $\deltah{k}$-m-reduction is also a $\deltah{k}$-1tt-reduction. We will see Theorem \ref{prop:1tt-hard}(2) later. 

The key idea of Hartmanis \etalc~\cite{HLY86} is to find, for any given reduction $f$, a restriction of its domain on which $f$ is honest. The following lemma is a generalization of a technical part of their proof. For any reduction $F$ and any set $A$, the {\em restriction of $F$ on $A$} is the function defined as $F$ on any input in $A$ and {\lq\lq}undefined{\rq\rq} on any input not in $A$.  

\begin{lemma}\label{lemma:hartmanis} 
Let $j,k\in \nat^{+}$. Assume that
$A\not\in\subdeltaexph{\max \{ j,k \}}$ and $B\in\deltaexph{j}$. 
\begin{enumerate}\vs{-2}
\item If $A$ is $\deltah{k}$-m-reducible to $B$ via $f$, then there exists a set $C$ in $\deltah{k}$ such that the restriction of $f$ on $C$ is honest and $A\cap C$ and $\overline{A}\cap C$ are infinite and coinfinite.
\vs{-2}
\item  If $A$ is $\deltah{k}$-1tt-reducible to $B$, then there exists a set $C$ in $\deltah{k}$ such that $A\cap C$ is h-$\deltah{k}$-1tt-reducible to $B$ and $A\cap C$ and $\overline{A}\cap C$ are infinite and coinfinite.
\end{enumerate}
\end{lemma}

Before proving Lemma \ref{lemma:hartmanis}, we present the following  lemma. For any function $\alpha$ from $\Sigma^*\times\{0,1\}$ to $\{0,1\}$, the notation $\alpha_x$ denotes the function mapping  $\{0,1\}$ to $\{0,1\}$ defined by $\alpha_x(y) = \alpha(x,y)$ for all $y\in\{0,1\}$.

\begin{lemma}\label{lemma:one-side-honest}
Let $A,B\subseteq\Sigma^*$ and $j\in\nat^{+}$. Let $r\in\{k\mathrm{tt,btt,tt,T}\}$. The following three statements are equivalent.
\vs{-2}
\begin{enumerate}
\item $A$ is h-$\deltah{j}$-$r$-reducible to $B$. 
\vs{-2}
\item $A$ is $\deltah{j}$-$r$-reducible to $B$ via a reduction whose restriction on $A$ is componentwise honest. 
\vs{-2}
\item $A$ is $\deltah{j}$-$r$-reducible to $B$ via a reduction whose restriction on $\overline{A}$ is componentwise honest. 
\end{enumerate}
\end{lemma}

\begin{proof}
Obviously, (1) implies both (2) and (3). It thus suffices to show that (2) implies (1) since we can show the implication of (3) to (1) in a similar way. Here, we show only the case where $r=k\mathrm{tt}$. Now, assume that $A$ is $\deltah{j}$-$k$tt-reducible to $B$ via a certain reduction pair $(f,\alpha)$ with the condition that  the restriction of $f$ on $A$ is componentwise honest. Let $f_1,f_2, \ldots , f_k$ be the $k$  $\fdeltah{j}$-functions satisfying that $f(x)=\pair{f_1(x),f_2(x), \ldots , f_k(x)}$ for every $x\in\dom(f)$. Let $p$ be any polynomial such that $|x| \leq p(|f_i(x)|)$ for all $x \in A$ and for each $i\in[1,k]_{\integer}$. For simplicity, assume that $p(n)\geq n$ for any $n\in\nat$. To show (1), we define the new reduction pair $(g,\beta)$ as follows. Take an arbitrary string $x$ and let $n=|x|$. If $n \leq p(|f_i(x)|)$ for all $i\in[1,k]_{\integer}$, then define $g(x)$  to be $f(x)$ and ${\beta}_x$ to be ${\alpha}_x $. Otherwise, define $g(x)$ to be the $k$-tuple $\pair{x,\ldots,x}$ (\ie the $k$ tuple consisting of all $x$'s) and ${\beta}_x$ to be the constant 0 function (that is, $\beta_x(y)=0$ for all $y\in\{0,1\}$). It follows that $(g,\beta)$ $\deltah{j}$-$k$tt-reduces $A$ to $B$. Therefore, $A$ is $\deltah{j}$-$k$tt-reducible to $B$.
\end{proof}

Now, we are ready to prove Lemma \ref{lemma:hartmanis}.

\begin{proofof}{Lemma \ref{lemma:hartmanis}}
We show only the second claim. The proof for the first claim is similar except for the use of Lemma \ref{lemma:one-side-honest}. For any index $i\in\nat$, let $K_i$ be any fixed $\p$-m-complete set for $\sigmah{i}$. For simplicity, let $i=\max \{ j,k \}$. Assume that $A$ is $\deltah{k}$-1tt-reducible to $B$ via a reduction pair $(f,\alpha)$. Since $f$ is a $1$tt-reduction, we can define $f'(x)$ to be $set(f(x))$ for each input $x$. For each number $\ell\in\nat^{+}$, define the set $C_{\ell}=\{x\mid |x| \leq |f'(x)|^{\ell}\}$. Clearly, $C_{\ell}$ belongs to $\deltah{k}$ since $f$ is in $\fdeltah{k}$. Note that $A$ is infinite and coinfinite because $A\not\in\subdeltaexph{i}$.

Since $f'\in\fdeltah{k}$, there exists a polynomial-time deterministic oracle TM $N$ that computes $f$ with $K_{k-1}$ as an oracle. Similarly, we have an exponential-time deterministic oracle TM $M$ recognizing $B$ with $K_{j-1}$ as an oracle. Take a large enough number $d$ and assume that the running time of $N$ is at most $n^d+d$ and that of $M$ is at most $2^{n^d+d}$ for all inputs of length $n$ and all oracles. Now, we claim the following. 
 
\begin{claim}
The sets $A\cap C_{\ell}$ and $\overline{A}\cap C_{\ell}$ are both infinite and coinfinite for all but finitely many numbers ${\ell}$ in $\nat^{+}$.
\end{claim}

The desired set $C$ is defined as $C_{\ell}$ for any fixed $\ell$ that satisfies the above claim. The claim guarantees that the restriction of $f'$ on $A\cap C$ is (componentwise) honest. Lemma \ref{lemma:one-side-honest} implies that $A\cap C_{\ell}$ is h-$\deltah{k}$-1tt-reducible to $B$.  

It still remains to prove the above claim. We only prove the claim for $A\cap C_{\ell}$. First, note that $A\cap C_{\ell}$ is always coinfinite since $\overline{A\cap C_{\ell}}$ contains $\overline{A}$. Next, we show that $A\cap C_{\ell}$ is infinite for all but finitely many numbers ${\ell}$. Assume to the contrary that there are infinitely many numbers ${\ell}>0$ for which $|x|^{1/{\ell}}>|f'(x)|$ for all but finitely many strings $x$ in $A$. Let $L$ be the collection of all such $\ell$'s. Take any element $\ell$ from $L$ and fix a number $n_{\ell}$ satisfying that, for all strings $x\in\Sigma^{\geq n_{\ell}}$, $|x|\leq |f'(x)|^{\ell}$ implies $x\not\in A$. Consider the following algorithm $\AAA$: 
\begin{quote}
{\sf On input $x$, let $n=|x|$. If $|x|\leq n_{\ell}$,
then output $A(x)$. Otherwise, first compute $f'(x)$ by running $N^{K_{k-1}}$ on input $x$. If $|x|\leq |f'(x)|^{\ell}$, then reject $x$. If $|x|>|f'(x)|^{\ell}$, then compute $z=M^{K_{j-1}}(f'(x))$ and output the value $\alpha_{x}(z)$.}
\end{quote}
The running time of $\AAA$ on the input $x$ is at most $c(|x|^d+d+2^{|x|^{d/\ell}+d})$ for a certain constant $c$ independent of the input. It thus follows that $\AAA$ recognizes $A$
in time $O(2^{n^{d/{\ell}}})$ with access to oracle $K_{i-1}$. This means that $A$ belongs to $\mathrm{DTIME}^{K_{i-1}}(2^{n^{d/{\ell}}})$. Since $\ell$ is arbitrary in $L$, $A$ belongs to $\bigcap_{\ell\in L} \mathrm{DTIME}^{K_{i-1}}(2^{n^{d/{\ell}}})$, which actually equals $\subdeltaexph{i}$ because $d$ is independent of $\ell$ and $L$ is infinite. This contradicts our assumption that $A\not\in\subdeltaexph{i}$. This complete the proof of the claim.
\end{proofof}

The original result of Hartmanis \etalc~\cite{HLY86} refers to $\p$-m-completeness of $\np$-simple sets. Recently, Schaefer and Fenner \cite{SF98} showed a similar result for $\p$-1tt-completeness. They proved that no $\np$-simple set is $\p$-1tt-complete for $\np$ if $\up \nsubseteq\subexp$. A key to their proof is the fact that $Sep(\subexp,\np)$ implies $\up\subseteq\subexp$, where $Sep(\CC,\DD)$ means the separation property of {Yamakami} \cite{Yam92} that, for any two disjoint sets $A,B\in \DD$, there exists a set $S$ in $\CC\cap\co\CC$  satisfying that $A\subseteq S\subseteq \overline{B}$. 

The following theorem shows
that the assumption {\lq\lq}$\up\nsubseteq\subexp${\rq\rq} of Schaefer and Fenner can be replaced by {\lq\lq}$\unique(\np\cap\co\np)\nsubseteq\subexp$.{\rq\rq} Note that $\up\subseteq\unique(\np\cap\co\np)$ since $\p\subseteq\np\cap\co\np$.
{}From Yamakami's \cite{Yam92} observation, $Sep(\subexp,\unique(\np\cap\co\np))$ holds if and only if $\unique(\np\cap\co\np)\subseteq\subexp$. 

\begin{theorem}\label{prop:1tt-hard} 
Let $j,k\in\nat^{+}$. 
\vs{-2}
\begin{enumerate}
\item No $\sigmah{k}$-immune set in $\deltaexph{j}$ is $\deltah{k}$-1tt-hard for $\unique(\sigmah{k}\cap\pih{k})$ if $\unique(\sigmah{k} \cap \pih{k}) \nsubseteq \subdeltaexph{\max\{j,k\}}$.
\vs{-2}
\item No $\sigmah{k}$-simple set is $\deltah{k}$-1tt-complete for $\sigmah{k}$ if $\unique(\sigmah{k} \cap \pih{k}) \nsubseteq \subdeltaexph{k}$.
\end{enumerate}
\end{theorem}

Our proof of Theorem \ref{prop:1tt-hard} follows from Lemmas \ref{lemma:hartmanis}, \ref{lemma:one-side-honest}, and \ref{lemma:1tt-complete}. 

\begin{lemma}\label{lemma:1tt-complete}
Let $k\in\nat^{+}$ and assume that $A$ is h-$\deltah{k}$-1tt-reducible to a $\sigmah{k}$-immune set $B$.
\vs{-2}
\begin{enumerate}
\item If $A$ belongs to $\sigmah{k}$, then there exists a set $C\in\deltah{k}$ and a total function $f\in\fdeltah{k}$ such that $\overline{A}\cap\overline{C}$ is finite, $f$ $\deltah{k}$-m-reduces $A$ to $\overline{B}$, and the restriction of $f$ on $C$ is honest. 
\vs{-2}
\item $A$ belongs to $\sigmah{k} \cap \pih{k}$ if and only if $A$ belongs to $\deltah{k}$.
\end{enumerate}
\end{lemma}

In other words, Lemma \ref{lemma:1tt-complete}(2) states that $\deltah{k}=\{A\in\sigmah{k}\cap\pih{k}\mid \exists B:\text{$\sigmah{k}$-immune}\ (A\honettreduces{\deltah{k}}B)\}$.

\begin{proofof}{Lemma \ref{lemma:1tt-complete}} 
Assume that $A$ is in $\sigmah{k}$, $B$ is $\sigmah{k}$-immune, and $A$ is h-$\deltah{k}$-1tt-reducible to $B$ via a reduction pair $(f,\alpha)$, where $f$ maps $\Sigma^*$ to $\Sigma^*$ and $\alpha$ maps $\Sigma^*\times\{0,1\}$ to $\{0,1\}$. 

1) Suppose that $A$ is in $\sigmah{k}$. Since $f(x)$ is always of the form $\pair{y}_1$, for brevity, we identify $\pair{y}_1$ with $y$ itself. With this identification, we simply write $y=f(x)$ instead of $\pair{y}_1=f(x)$. It thus follows that $A=\{x\mid \alpha_x (B(f(x)))=1\}$. For convenience, let $\mathrm{ONE}$ and $\mathrm{FLIP}$ be the functions from $\{0,1\}$ to $\{0,1\}$ defined as  $\mathrm{ONE}(y)=1$ and $\mathrm{FLIP} (y)= 1 -y$ for all $y\in\{0,1\}$.

First, we want to show the existence of a number $n_0\in\nat$ satisfying the following condition: for any string $x\in A$ with $|x|\geq n_0$, ${\alpha}_x$ becomes either $\mathrm{ONE}$ or $\mathrm{FLIP}$. Consider the set
$A_0= \{ x \in A \mid \alpha_x (0)= 0 \}$. It follows that $f(A_0)\subseteq B$ because $y=f(x)$ and $x\in A_0$ imply $\alpha_x(B(y))\neq \alpha_x(0)$, which leads to $B(y)=1$. 
Since $f$ is honest, the set $f(A_0)$ belongs to $\sigmah{k}$. The $\sigmah{k}$-immunity of $B$ requires $f(A_0)$ to be finite and thus, $A_0$ is finite because of the honesty condition on $f$. The desired number $n_0$ can be defined as the minimal number such that $|x|<n_0$ for all $x\in A_0$.

Using this $n_0$, $A$ can be expressed as the union of the following three disjoint sets: $A_1 = A \cap
\{ 0,1 \}^{< n_0}$, $A_2 =
\{ x \mid |x| \geq n_0\:\wedge\: {\alpha}_x = \mathrm{ONE} 
\}$, and  $ A_3 = 
\{ x \mid |x| \geq n_0\:\wedge\: {\alpha}_x = \mathrm{FLIP} 
 \:\wedge\: f(x) \not\in B \}$. 
Obviously, $A_1$ is finite and $A_2$ is in $\deltah{k}$ because  $\alpha$ is in $\fdeltah{k}$. Now, define $C=\Sigma^{\geq n_0}\cap \overline{A_2}$. Note that $\overline{C}=\Sigma^{<n_0}\cup A_2$.
Clearly, $C$ is in $\deltah{k}$ since so is $A_2$. Note that $\overline{A}\cap \overline{C}$ is finite because $\overline{A}\cap \overline{C}\subseteq \Sigma^{<n_0}$. Now, we explicitly define the $\deltah{k}$-m-reduction $g$ from $A$ to $\overline{B}$ in the following way. Let $z_0$ and $z_1$ be any two fixed strings such that $z_0\not\in B$ and $z_1\in B$. Let $g(x)=z_0$ if $x\in A_1\cup A_2$; $g(x)=z_1$ if $x\in \overline{A_1}\cap\Sigma^{<n_0}$; $g(x)=f(x)$ if $x\in C$. The restriction $g$ on $C$ is honest because $f$ is honest. 
It is clear that $g$ $\deltah{k}$-m-reduces $A$ to $\overline{B}$.

2) It suffices to show the {\lq\lq}only if{\rq\rq} part. Assume that $A$ belongs to $\sigmah{k} \cap \pih{k}$. We have already proven in 1) the existence of a number $n_0$ such that $\alpha_x$ is either $\mathrm{ONE}$ or $\mathrm{FLIP}$ for any $x\in A$ with $|x|\geq n_0$. For this $n_0$, consider the set $C = 
\{ x \mid |x| \geq n_0 \:\wedge\: {\alpha}_x = \mathrm{FLIP} \:\wedge\:
 f(x) \in B \}$. Note that $C$ equals $\{ x \mid |x| \geq n_0 \:\wedge\: {\alpha}_x = \mathrm{FLIP} \:\wedge\:  x \not\in A \}$. 
Since $A \in \pih{k}$ and $\alpha\in\fdeltah{k}$, $C$ belongs to $\sigmah{k}$. Recall the definition of the set $A_3$ from 1). Since $B$ is $\sigmah{k}$-immune and $f$ is honest, $C$ should be finite. Thus, for all but finitely many strings $x$, ${\alpha}_x = \mathrm{FLIP} $ if and only if $x\in A_3$. Therefore,
$A_3$ belongs to $\deltah{k}$ and, as a consequence, $A$ belongs to $\deltah{k}$.
\end{proofof}

Finally, we present the proof of Theorem \ref{prop:1tt-hard}. Note that $\unique(\sigmah{k}\cap\pih{k})\subseteq\sigmah{k}$ at any level $k\in\nat^{+}$; see Yamakami \cite{Yam92}.

\begin{proofof}{Theorem \ref{prop:1tt-hard}}
We show only the first claim because the second claim follows from the first one. Now, take arbitrary numbers $j,k\in\nat^{+}$ and let $i=\max\{j,k\}$. Assume that $B$ is $\sigmah{k}$-immune and in $\deltaexph{j}$ and that $B$ is $\deltah{k}$-1tt-hard for $\unique(\sigmah{k}\cap\pih{k})$. Moreover, assume that there exists a set $A$ in $\unique(\sigmah{k}\cap\pih{k})$ but not in $\subdeltaexph{i}$. We want to derive a contradiction from these assumptions. 

Since $A\in\unique(\sigmah{k}\cap\pih{k})$, $A$ is of the form $\{x\mid \exists y[\pair{x,y}\in Graph(f)]\}$ for a certain polynomially-bounded single-valued partial function $f$ whose graph is in $\sigmah{k}\cap\pih{k}$. Take a polynomial $p$ such that $|f(x)|\leq p(|x|)$ for all $x\in\dom(f)$. Similar to Yamakami's \cite{Yam92} argument, we define $A_1=\{ \pair{x,z}\mid \exists y\in\Sigma^{\leq p(|x|)}[z\leq y\:\wedge\: \pair{x,y}\in Graph(f)]\}$ and $A_2=\{ \pair{x,z}\mid \exists y\in\Sigma^{\leq p(|x|)}[z>y\:\wedge\: \pair{x,y}\in Graph(f)]\}$. Clearly, $A_1$ and $A_2$ are mutually disjoint and in $\unique(\sigmah{k}\cap\pih{k})\subseteq\sigmah{k}$. Notice that $A\onereduces{\p}A_1$ and $A\onereduces{\p}A_2$. Hence, neither of $A_1$ and $A_2$ belong to $\subdeltaexph{i}$. Because of the $\deltah{k}$-1tt-hardness of $B$, $A_1$ is $\deltah{k}$-1tt-reducible to $B$. By Lemmas \ref{lemma:one-side-honest} and \ref{lemma:hartmanis}, we can take a set $C$ in $\deltah{k}$ such that $A_1\cap C$ is infinite and coinfinite and $A_1\cap C$ is h-$\deltah{k}$-1tt-reducible to $B$. Moreover, since $A_1\cap C$ is in $\sigmah{k}$, by Lemma \ref{lemma:1tt-complete}(1), there exist a set $D\in\deltah{k}$ and a total $\fdeltah{k}$-function $g$ such that $\overline{A_1\cap C}\cap \overline{D}$ is finite, $g$ $\deltah{k}$-m-reduces $A_1\cap C$ to $\overline{B}$, and the restriction of $g$ on $D$ is honest. Since $A_2\cap D$ is a subset of $\overline{A_1\cap C}\cap D$, the image $g(A_2\cap D)$ is in $B$.  Moreover, $A_2\cap D$ is infinite because, otherwise, from the inclusion $A_2\cap \overline{D}\subseteq \overline{A_1\cap C}\cap\overline{D}$, $A_2$ is finite. Since $g$ is honest on the domain $D$, $g(A_2\cap D)$ must be infinite. Since $g(A_2\cap D)$ is in $\sigmah{k}$, $B$ cannot be $\sigmah{k}$-immune, a contradiction.
\end{proofof}

A natural open question arising from Theorem \ref{prop:1tt-hard} is to determine whether the assumption $\unique(\sigmah{k}\cap\pih{k})$ used in the theorem is practically optimal in such a sense that there exists a relativized world in which $\sigmah{k}\nsubseteq\subdeltaexph{k}$ but a $\sigmah{k}$-simple $\deltah{k}$-1tt-complete set for $\sigmah{k}$ exists.

\section{Epilogue}\label{sec:conclusion}

The class $\np$ and the polynomial hierarchy built over $\np$ have been a major subject in the theory of computational complexity since their introduction. Since an early work of Flajolet and Steyaert \cite{FS74}, the notions of resource-bounded immunity and simplicity have made significant contributions to the development of the theory. In this paper, we have further explored these notions for a better understanding of the classes lying in the polynomial  hierarchy. Although our research is foundational in nature, we hope that our research will draw general-audience's attention to the importance of resource-bounded immunity and simplicity. There are, of course, many open problems waiting to be solved. 
For the future research, we list six important directions on the study of resource-bounded immunity and simplicity.

\begin{enumerate}\vs{-2}
\item {\sf Separating weak notions of completeness.} The completeness notions in general have played an essential role in computational complexity theory for the past three decades. Recall that Post originally introduced immunity and simplicity in an attempt to fill the gap between the class of recursive sets and the class of complete sets. One of the open problems is the differentiation of weak completeness within the polynomial hierarchy. We hope that weak completeness notions can be separated by different types of simplicity notions. 
\vs{-2}
\item {\sf Finding new connections to other notions.} Although we have mentioned a close connection of immunity to several notions, such as complexity cores and instance complexity, we know few connections to other existing complexity-theoretical notions. We still need to discover new connections to other important notions; for instance, resource-bounded measure and pseudorandom generators. 
\vs{-2}
\item {\sf Studying the nature of collections of immune sets.} We have shown several closure properties of the classes of certain types of immune sets under weak reductions. For a better understanding of immunity, it is also important to study {\em classes} of immune sets rather than each individual immune set alone. There are few systematic studies along this line. We hope to discover useful closure properties that are unique to these classes of immune sets. Such properties may reveal the characteristics of immune sets.
\vs{-2}
\item {\sf Exploring limited immunity and simplicity notions.} We have reopened the study of $k$-immunity and further introduced the notion of feasibly $k$-immunity to analyze the structure of $\np$. An open question is to determine whether there exists a feasible $k$-simple set. Along a similar line of research, we can naturally introduce strongly $k$-immunity, almost $k$-immunity, and $k$-hyperimmunity. We hope to explore these notions and study their roles within $\np$.  
\vs{-2}
\item {\sf Exploring new relativized worlds.} An oracle separation by immune or simple sets is sometimes called a {\em strong separation}. Such a separation usually requires intricate tools and proof techniques. We need to develop new tools and techniques to exhibit tight separations of complexity classes. An important open problem, for instance, is to find a relativized world where  a $\sigmah{k}\cap\pih{k}$-immune $\sigmah{k}$-simple set exists for every $k\geq2$. 
\vs{-2}
\item {\sf Discovering good working hypotheses.} We have proposed the k-immune hypothesis, which implies the existence of an $\np$-simple set. We hope that finding a good working hypothesis will boost the research on immunity and simplicity. 
\end{enumerate}


\end{document}